\documentclass[pdflatex,sn-mathphys-num]{sn-jnl}

\usepackage{graphicx}%
\usepackage{multirow}%
\usepackage{amsmath,amssymb,amsfonts}%
\usepackage{amsthm}%
\usepackage{mathrsfs}%
\usepackage[title]{appendix}%
\usepackage{xcolor}%
\usepackage{textcomp}%
\usepackage{manyfoot}%
\usepackage{booktabs}%
\usepackage{listings}%

\usepackage{dcolumn}
\usepackage{bm}
\usepackage{multirow}
\usepackage{amssymb}
\usepackage{hyperref}
\usepackage{subcaption}
\usepackage{float}
\usepackage{makecell}
\usepackage[section]{placeins}

\theoremstyle{thmstyleone}%
\theoremstyle{thmstyletwo}%

\theoremstyle{thmstylethree}%

\begin{document}

\title[Article Title]{CFD-driven symbolic regression of RANS transition models}


\author*[1]{\fnm{Louenas} \sur{Zemmour}}\email{louenas.zemmour@sorbonne-universite.fr}

\author[2]{\fnm{Xavier} \sur{Gloerfelt}}\email{xavier.gloerfelt@ensam.eu}

\author[1]{\fnm{Paola} \sur{Cinnella}}\email{paola.cinnella@sorbonne-universite.fr}

\affil*[1]{\orgdiv{Sorbonne Université, CNRS}, \orgname{Institut Jean Le Rond D'Alembert}, \orgaddress{\street{4 Pl. Jussieu}, \city{Paris}, \postcode{75005}, \country{France}}}

\affil[2]{\orgdiv{DynFluid Laboratory}, \orgname{Arts et Metiers Institute of Technology}, \orgaddress{\street{151 Bd. de l'H\^opital}, \city{Paris}, \postcode{75013}, \country{France}}}

\abstract{\unboldmath Laminar--turbulent transition governs skin friction, heat transfer and losses in turbomachinery, yet Reynolds-averaged Navier--Stokes (RANS) transition models rely on empirical correlations and hand-calibrated activation functions whose structural deficiencies cannot be resolved by recalibration. We propose a CFD-driven sparse symbolic regression framework to learn interpretable, closed-form corrections to the production and destruction activation functions in the intermittency equation of the local, Galilean-invariant $k$--$\omega$--$\gamma$ model of Ge \textit{et al.}~\cite{ge2014bypass}. Each correction combines a polynomial of a local pressure-gradient parameter, controlling amplitude and sign, with a bounded exponential of local flow features, controlling spatial support. This construction bounds the corrections without ad hoc limiters. Coefficients are identified by gradient-free optimization with the RANS solver in the loop, accelerated by POD--Kriging field surrogates, using mean-velocity data from wall-resolved large-eddy simulations (LES) of the ERCOFTAC T3 flat-plate cases. Elastic-net hyperparameters are selected automatically through bi-level optimization with leave-one-case-out cross-validation. The resulting model retains 11 of 21 candidate terms, compared with 20 under conventional grid-search tuning, while achieving comparable or slightly better accuracy on training and unseen cases. Both approaches recover the same leading-order correction. Applied without retraining to the transonic LS89 turbine cascade, far beyond the training envelope, the augmented model remains numerically robust and improves heat-transfer predictions at all four operating points, reducing errors by up to $70\%$ where the baseline predicts premature transition under high freestream turbulence. Residual errors in transition onset and length highlight the limits of training on flat-plate data alone.}\keywords{Laminar--turbulent transition, RANS modeling, Intermittency, Symbolic regression, CFD-driven learning, Bi-level optimization, Turbomachinery}

\maketitle

\section{Introduction}\label{sec:introduction}
Transitional flows are ubiquitous in turbomachinery, wind turbines and aircraft, where laminar-to-turbulent transition governs skin friction, heat transfer and flow
separation, and hence performance and thermal loads. In turbomachinery, blades are exposed to high freestream turbulence levels and periodic wakes, and bypass
transition is a dominant mechanism~\cite{mayle1991role, durbin2017perspectives}. Transition prediction nevertheless remains a challenge for computational fluid
dynamics (CFD), owing to its sensitivity to a large set of parameters, including the Reynolds and Mach numbers, the freestream turbulence intensity and length scale, pressure gradients, surface roughness
and three-dimensional effects~\cite{mayle1991role, arnal2000laminar}.
Scale-resolving simulations have clarified the mechanisms of bypass transition on flat plates~\cite{jacobs2001simulations, brandt2004transition, zaki2005mode,
ovchinnikov2008numerical} and in turbomachinery passages~\cite{zaki2010direct, zhao2020bypass, dupuy2020analysis}, and reproduce experiments accurately provided that
the freestream disturbances are carefully prescribed~\cite{pinto2019synthetic, bienner2024influence}. They remain, however, too costly for routine design, so that
transition models for Reynolds-averaged Navier--Stokes (RANS) solvers are still an active research topic. To be usable in general-purpose solvers, such models should
rely on local quantities only, so as to remain compatible with unstructured grids and massively parallel computations.

Early methods based on linear stability theory~\cite{smith1956transition, van1956suggested} predict transition when the logarithm $N$ of the amplification factor of
the most unstable disturbances reaches an empirical threshold. These $e^N$ methods have been successful in applied aerodynamics~\cite{vaningen2008eN} and have
been made local by transporting the amplification factor~\cite{coder2014computational, coder2017enhancement}. They describe, however, the modal growth of
Tollmien--Schlichting waves, the freestream turbulence entering only through empirical adjustments of the critical $N$ factor, valid at low turbulence
levels~\cite{mack1977transition}. They are therefore unsuitable for bypass transition, in which the non-modal growth of streaks forced by high-amplitude freestream
disturbances bypasses the modal route~\cite{durbin2017perspectives, dick2017transition}.

Intermittency-based models describe transition through the intermittency factor, i.e.\ the fraction of time during which the flow is turbulent~\cite{dhawan1958some}.
Steelant \& Dick~\cite{steelant1996modelling} coupled an intermittency transport equation with conditioned Navier--Stokes equations, i.e.\ separate sets of
conditionally averaged equations for the laminar and turbulent parts of the flow, coupled through the intermittency. Solving two sets of flow equations is, however,
expensive and incompatible with general-purpose solvers, designed for a single set of RANS equations. Suzen \& Huang~\cite{suzen2000modeling} therefore adopted a
single-set formulation, in which the intermittency multiplies the eddy viscosity, but their model still relies on non-local quantities, namely onset correlations
based on boundary-layer-edge values and the streamwise distance from the onset location. Evaluating such quantities requires searches along grid lines or streamlines,
which are impractical on complex geometries, especially on unstructured meshes and in parallel computations.

Considerable efforts have since been devoted to models based solely on local quantities. The $k$--$\omega$--$\gamma$--$Re_{\theta t}$ model of Menter \emph{et al.}~\cite{menter2006transition, langtry2009correlation} transports the intermittency and a transition-onset Reynolds number, the local momentum-thickness
Reynolds number being estimated from the vorticity Reynolds number~\cite{van_driest}. Its correlations, initially not released, were recalibrated by several
authors~\cite{piotrowski2008transition, suluksna2008assessment, suluksna2009correlations, content2010application} before being published~\cite{langtry2009correlation};
the model has since been widely applied to turbomachinery flows~\cite{benyahia2011transition, benyahia2011separation}, but proved sensitive to numerical schemes and
implementation details~\cite{cheng2009validation}. Moreover, early local formulations such as the $\gamma$--$Re_{\theta t}$ model involve the local velocity magnitude,
e.g.\ in the estimates of the freestream turbulence intensity and of the pressure-gradient parameter, and are therefore not Galilean invariant~\cite{menter2015one}.
This motivated the simpler, Galilean-invariant one-equation $\gamma$ model of Menter \textit{et al.}~\cite{menter2015one}. Being calibrated on limited data,
correlation-based models may fail outside their calibration range, as reported for the high-Reynolds-number flows of wind turbines~\cite{colonia2017assessment}, and
their predictions of transonic turbine flows are highly sensitive to the uncertainties of the correlations~\cite{pecnik2013assessment}. Other models build the
intermittency source terms from local flow sensors, either algebraically~\cite{kubacki2016algebraic} or through a transport equation~\cite{durbin2012intermittency,
ge2014bypass}. In the latter, used here as baseline, transition results from the diffusion of intermittency into the boundary layer and is controlled by ramp
functions of the vorticity Reynolds number and of the eddy-viscosity ratio.

Laminar kinetic energy (LKE) models, originating from the work of Mayle \& Schulz~\cite{mayle1997heat}, instead transport the energy $k_L$ of the pre-transitional
fluctuations, which is transferred to the turbulent kinetic energy as transition proceeds~\cite{walters2004new, walters2005computational}. Walters \&
Cokljat~\cite{walters2008three} refined this approach in a $k$--$\omega$ framework and assessed it on the ERCOFTAC T3 flat plates~\cite{savill1993evaluating} and on
airfoils. Juntasaro \& Ngiamsoongnirn~\cite{juntasaro2014new} combined both families by defining $\gamma = k/(k+k_L)$, whose transport equation is derived from the
$k$-equation of the SST model~\cite{menter1994two} and the $k_L$-equation of Walters \& Cokljat, and later extended their model to pressure
gradients~\cite{juntasaro2017gamma}. Comparative assessments nevertheless show that no model is uniformly successful~\cite{cutrone2007evaluation, dick2017transition}:
in their review of turbomachinery applications, Dick \& Kubacki~\cite{dick2017transition} found that the $\gamma$--$Re_{\theta t}$ model performs well overall but
requires improvement for transition in separated flows, whereas the $k$--$k_L$--$\omega$ LKE model of Walters \& Cokljat fails to predict wake-induced transition.

All these models share a common limitation: their functional forms and closure coefficients were devised from physical reasoning and human expertise, and calibrated
by hand on limited experimental data, mostly flat-plate boundary layers. They are thus data-driven in an empirical way, but lack a systematic procedure for
consistently formulating and calibrating them from data. They consequently remain tailored to specific transition mechanisms, and recent developments have mostly
consisted of case-specific recalibrations~\cite{colonia2017assessment} rather than corrections of their structural deficiencies.

Machine learning (ML) has recently emerged as a means of improving RANS closures from high-fidelity data~\cite{duraisamy2019turbulence, xiao2019quantification,
duraisamy2021perspectives, PaolaVKI, Recentreview}. In \textit{a priori}, or frozen-field, approaches, corrections computed from high-fidelity fields are regressed onto
flow features, e.g.\ with neural networks~\cite{ling2016reynolds}, gene expression programming (GEP)~\cite{weatheritt2016novel} or sparse
regression~\cite{schmelzer2020discovery, cherroud2022sparse, cinnellaSR}. They are inexpensive, but accuracy on the training targets does not guarantee accurate
predictions once the corrections are coupled with the solver, because the RANS equations may be ill-conditioned with respect to the learned
terms~\cite{wu2019ill, duraisamy2021perspectives}. In field inversion and machine learning (FIML)~\cite{parish2016paradigm}, a correction field is first inferred with
the RANS solver, usually through an adjoint method, and then generalized as a function of local features. In CFD-driven, or model-consistent, approaches, the solver is
embedded in the training loop, candidate models being assessed on converged RANS solutions~\cite{holland2019field, zhao2020rans, lav2021rans, 
strofer2021end, saidi2022cfd}, and the cost can be mitigated by surrogate models~\cite{saidi2022cfd, rincon2023progressive, fang2025surrogate}.

For transition modeling, Duraisamy \textit{et al.}~\cite{duraisamy2015new} and Srivastava \& Duraisamy~\cite{srivastava2021generalizable} used FIML to correct
intermittency-based models; the latter generalized to turbine cascades, but required an adjoint solver and predicted premature transition for combinations of pressure
gradient and turbulence intensity absent from the training data. Yang \& Xiao~\cite{yang2020improving} applied FIML to a four-equation transition model, using an
ensemble Kalman filter instead of an adjoint. Wu \textit{et al.}~\cite{wu2022two} and Li \textit{et al.}~\cite{li2023machine} predicted the intermittency or the
eddy viscosity with neural networks trained \textit{a priori}, and Zafar \textit{et al.}~\cite{zafar2020convolutional} used neural networks to predict amplification
factors in the framework of linear stability theory, at the expense of interpretability and extrapolation capability. Interpretable corrections of LKE models for
turbine flows were obtained with CFD-driven GEP~\cite{akolekar2021transition, akolekar2022multi, fang2024data}, leading to the reformulated model of Pacciani
\textit{et al.}~\cite{pacciani2025reformulation}; GEP is, however, stochastic, and some of the features used were not Galilean invariant. Simoni
\textit{et al.}~\cite{simoni_rans} used the deterministic elastic-net regression, but trained their LKE model \textit{a priori} on an experimental database.

CFD-driven learning is particularly well suited to transition modeling for two reasons. First, intermittency is a modeling construct: its extraction from high-fidelity data depends on non-unique conditional-sampling procedures and thresholds~\cite{nolan2012conditional, veerasamy2020rational}, making \textit{a priori} targets for intermittency and its source terms ambiguous. Second, transition models are strongly nonlinear: improving intermittency \textit{a priori} may substantially degrade skin-friction predictions once the model is coupled with the RANS equations~\cite{capel2026decoding}. In the presence of model inadequacy, a model term must therefore be assessed through its effect on observable predictions, allowing its calibration to compensate for errors elsewhere in the model~\cite{kennedy2001bayesian}. In FIML, the corrective field inferred through data assimilation is regressed in a separate step, which does not guarantee consistency between the learned correction and the coupled solver~\cite{holland2019field}. Training corrections directly against RANS predictions of observable quantities addresses this issue, but entails a nonlinear, high-dimensional optimization problem with the solver in the loop, at substantial computational cost when adjoint gradients are unavailable. To date, CFD-driven learning of transition models has mainly relied on GEP~\cite{akolekar2021transition, fang2024data}, which typically requires many costly CFD evaluations.

In this work, we combine CFD-driven training with sparse symbolic regression to augment the intermittency transport equation of the $k$--$\omega$--$\gamma$ model of
Ge \textit{et al.}~\cite{ge2014bypass}. First, we learn additive corrections to the activation functions of the production and destruction terms, expressed as the
product of a polynomial of a local pressure-gradient parameter, which sets their amplitude and sign, and of an exponential of local, Galilean-invariant features,
which sets their spatial support. The exponential is bounded by construction, so that the augmented model preserves the structure of the baseline and can be
integrated in the RANS solver without ad hoc limiters. Second, building on the framework of Ben Hassan Sa\"idi \textit{et al.}~\cite{saidi2022cfd}, we train the
model coefficients through a gradient-free, surrogate-assisted optimization with the RANS solver in the loop. Elastic-net regularization promotes sparsity and
robustness, but its hyperparameters are conventionally tuned by cross-validating models trained on a grid of hyperparameter values, which is costly and whose outcome
depends on the arbitrary choice of the grid. To reduce this burden while exploring the hyperparameter space continuously, we introduce an automatic bi-level
optimization strategy combined with a leave-one-case-out cross-validation. The models are trained on large-eddy simulation (LES) data of
freestream-turbulence-induced transition on flat plates~\cite{zemmour2026high}, for a range of freestream turbulence intensities and pressure gradients. The LES
database mimics and complements the ERCOFTAC T3 experiments~\cite{roach1992influence, savill1993evaluating}, while providing a controlled numerical set-up and full
flow fields for data-driven model development. Finally, the most promising model is applied without retraining to the transonic LS89 turbine
cascade~\cite{arts1992aero}, whose curvature, acceleration, Mach and Reynolds numbers lie well outside the training data. This stress test delineates the benefits and
the limitations of the proposed approach.

The paper is organized as follows. Section~\ref{sec:governing_equations} presents the governing equations and the baseline transition
model, and Section~\ref{sec:rans_model_augmentation} the proposed corrections. Section~\ref{sec:database} describes the training and test data, and
Section~\ref{sec:cfd_driven_identification} the CFD-driven identification procedure under two hyperparameter optimization strategies. The identified models are
analyzed and assessed on flat-plate flows in Section~\ref{sec:results}, and applied to the LS89 cascade in Section~\ref{sec:ls89}. Conclusions and perspectives are
drawn in Section~\ref{sec:conclusions}.

\section{Baseline transition model}\label{sec:governing_equations}\label{sec:baseline_model}
The flow is governed by the compressible Reynolds-averaged Navier--Stokes (RANS) equations, written in conservative form for the mean density, velocity and total
energy, the molecular viscosity $\mu$ being given by Sutherland's law. The Reynolds stresses are modeled with the Boussinesq hypothesis,
$\tau_{ij} = 2 \mu_t (S_{ij} - S_{kk} \delta_{ij}/3) - 2 \rho k \delta_{ij}/3$, where $S_{ij} = (\partial u_i / \partial x_j + \partial u_j / \partial x_i)/2$ is
the mean strain-rate tensor, $k$ the turbulent kinetic energy and $\mu_t$ the eddy viscosity, and the turbulent heat flux with a Reynolds analogy based on a turbulent
Prandtl number. The eddy viscosity is provided by the $k$--$\omega$ model of Wilcox~\cite{wilcox1988reassessment}, in which the production of turbulent kinetic
energy is weighted by the intermittency $\gamma$, as proposed by Ge \textit{et al.}~\cite{ge2014bypass}. To reduce the stiffness associated with the steep near-wall
gradients of $\omega$ and to guarantee its positivity, the model is rewritten in terms of $\hat{\omega} = \ln \omega$~\cite{bassi2005discontinuous}:
\begin{align}
    \frac{\partial (\rho k)}{\partial t} + \frac{\partial (\rho u_j k)}{\partial x_j} &= \gamma P_k - C_{\mu} \rho k e^{\hat{\omega}}
    + \frac{\partial}{\partial x_j} \left[ \left( \mu + \sigma_k \mu_t \right) \frac{\partial k}{\partial x_j} \right], \label{eq:kom_k}\\
    \frac{\partial (\rho \hat{\omega})}{\partial t} + \frac{\partial (\rho u_j \hat{\omega})}{\partial x_j} &= C_{\omega 1} \frac{\mathcal{P}}{k} - C_{\omega 2} \rho e^{\hat{\omega}}
    + \frac{\partial}{\partial x_j} \left[ \left( \mu + \sigma_{\omega} \mu_t \right) \frac{\partial \hat{\omega}}{\partial x_j} \right] \nonumber \\
    &\quad + \left( \mu + \sigma_{\omega} \mu_t \right) \frac{\partial \hat{\omega}}{\partial x_j} \frac{\partial \hat{\omega}}{\partial x_j}, \label{eq:kom_w}
\end{align}
with $\mu_t = \rho k e^{-\hat{\omega}}$. The last term of Eq.~\eqref{eq:kom_w} results from the change of variable. The production of turbulent kinetic energy by
the mean flow, $\mathcal{P} = \tau_{ij} \, \partial u_i / \partial x_j$, is bounded in Eq.~\eqref{eq:kom_k} by the realizability limiter adopted by Ge
\textit{et al.}~\cite{ge2014bypass}:
\begin{equation}\label{eq:production_limiter}
    P_k = \min \left( \mathcal{P}, \frac{\rho k |S|}{\sqrt{3}} \right), \qquad |S| = \sqrt{S^*_{ij} S^*_{ij}}, \qquad S^*_{ij} = S_{ij} - \frac{1}{3} S_{kk} \delta_{ij}.
\end{equation}
The model constants are $C_{\mu} = 0.09$, $C_{\omega 1} = 5/9$, $C_{\omega 2} = 3/40$ and $\sigma_k = \sigma_{\omega} = 0.5$~\cite{wilcox1988reassessment}.

The intermittency $\gamma$, which represents the fraction of time during which the flow is turbulent at a given location, tends to zero in laminar regions,
where the production of turbulent kinetic energy is thus suppressed, and is close to unity in fully turbulent regions, where the original $k$--$\omega$ model is
recovered. In the $k$--$\omega$--$\gamma$ model of Ge \textit{et al.}~\cite{ge2014bypass}, which extends the intermittency model of
Durbin~\cite{durbin2012intermittency} to general-purpose CFD codes, Eqs.~\eqref{eq:kom_k}--\eqref{eq:kom_w} are supplemented with a transport equation for the
intermittency, written in conservative form as
\begin{equation}\label{eq:intermittency}
    \frac{\partial (\rho \gamma)}{\partial t} + \frac{\partial (\rho u_j \gamma)}{\partial x_j} = \frac{\partial}{\partial x_j} \left[ \left( \frac{\mu}{\sigma_L}
    + \frac{\mu_t}{\sigma_{\gamma}} \right) \frac{\partial \gamma}{\partial x_j} \right] + P_{\gamma} - E_{\gamma},
\end{equation}
with $\sigma_L = 5$ and $\sigma_{\gamma} = 0.2$, $\gamma = 1$ in the freestream and $\partial \gamma / \partial n = 0$ at walls. In this model, bypass transition results
from the diffusion of intermittency from the freestream into the boundary layer: once $\gamma$ becomes non-zero, turbulent kinetic energy is produced, which increases
the eddy viscosity and further enhances the diffusion of $\gamma$. The production and destruction terms read
\begin{equation}\label{eq:source_terms}
    P_{\gamma} = F_{\gamma} \rho \Omega (\gamma_{\mathrm{max}} - \gamma) \sqrt{\gamma}, \qquad
    E_{\gamma} = G_{\gamma} F_{\mathrm{turb}} \rho \Omega \gamma^{1.5},
\end{equation}
where $\gamma_{\mathrm{max}} = 1.1$, $\Omega = \sqrt{2 W_{ij} W_{ij}}$ is the vorticity magnitude and $W_{ij} = \frac{1}{2} \left( \partial u_i / \partial x_j
- \partial u_j / \partial x_i \right)$ the rotation-rate tensor. The destruction term keeps the boundary layer laminar upstream of transition, and is switched off
in turbulent regions by the function
\begin{equation}\label{eq:turbulence_factor}
    F_{\mathrm{turb}} = \exp[-(Re_v R_t)^{1.2}]
\end{equation}
where $R_t = \mu_t/\mu$ is the eddy-viscosity ratio and $Re_v = \rho d_w^2 \Omega/(2.188 \mu)$ the vorticity Reynolds number, $d_w$ being the distance to the
nearest wall. The vorticity Reynolds number $Re_v$ is a local surrogate of the momentum-thickness Reynolds number, to which its maximum across a Blasius boundary
layer is equal~\cite{van_driest, menter2006transition, ge2014bypass}. Transition onset is controlled by a critical Reynolds number,
\begin{equation}\label{eq:critical_reynolds}
    Re_c = 400 - 360 \min \left( \frac{T_{\omega}}{2}, 1 \right),
\end{equation}
where $T_{\omega} = R_t\, \Omega/\omega$ is a sensor for the freestream turbulence. The critical Reynolds number decreases as $T_{\omega}$ increases, thus leading
to earlier transition at higher turbulence levels.

The activation functions $F_{\gamma}$ and $G_{\gamma}$ are linear ramps comparing the local vorticity Reynolds number with threshold values:
\begin{align}
    F_{\gamma} &= 2 \max \left[ 0, \min \left( 100 - 0.7 Re_v, 1 \right) \right] \times \min \left[ \max \left( Re_v - Re_c, 0 \right), 4 \right], \label{eq:ramp_production}\\
    G_{\gamma} &= 7.5 \max \left[ 0, \min \left( 100 - Re_v, 1 \right) \right] \times \min \left[ \max \left( Re_v - 18, 0 \right), 1 \right]. \label{eq:ramp_destruction}
\end{align}
In both functions, the second factor activates the source term where $Re_v$ exceeds a threshold value, namely $Re_c$ for the production and 18 for the
destruction, whereas the first factor switches it off where $Re_v$ exceeds about 143 and 100, respectively. The correction for separation-induced transition
proposed in~\cite{ge2014bypass}, which allows the intermittency to exceed unity in separated shear layers, is not used in this work.

The RANS equations and the turbulence and transition models are solved with the in-house code \textit{MUSICAA}, in which all transport equations are discretized in
conservative form on structured multi-block curvilinear grids with high-order centered finite differences~\cite{bienner2024multiblock}.

\section{Transition model augmentation}\label{sec:rans_model_augmentation}
The $k$--$\omega$--$\gamma$ model relies on empirical source terms, heuristic activation functions and calibrated closure coefficients. In the following,
we introduce a methodology to augment such formulations with data-driven analytical corrections. More specifically, we augment the activation
functions $F_{\gamma}$ and $G_{\gamma}$, which drive the production and destruction terms of the intermittency equation~\eqref{eq:intermittency},
with additive corrective terms:
\begin{align}
    \tilde{P}_{\gamma} &= \left( F_{\gamma} + R^P(\boldsymbol{\eta}) \right) \rho \Omega (\gamma_{\mathrm{max}} - \gamma) \sqrt{\gamma}, \label{eq:augmented_production}\\
    \tilde{E}_{\gamma} &= \left( G_{\gamma} + R^E(\boldsymbol{\eta}) \right) F_{\mathrm{turb}} \rho \Omega \gamma^{1.5}. \label{eq:augmented_destruction}
\end{align}
The corrections $R^P$ and $R^E$ account for the structural inadequacy of the ramp functions $F_{\gamma}$ and $G_{\gamma}$, which depend only on the vorticity
Reynolds number and on the critical Reynolds number, and are expressed as functions of a vector of local flow features
$\boldsymbol{\eta} = (\eta_0, \eta_1, \eta_2, \eta_3)$, defined in Section~\ref{sec:features}. Correcting the activation functions, rather than adding independent
source terms to the transport equation, preserves the structure of the baseline model, and the dependence of the source terms on $\gamma$.

The relation between the features $\boldsymbol{\eta}$ and the corrections $R^P$ and $R^E$ is learned by symbolic regression~\cite{cinnellaSR}, and more
specifically by a sparse identification algorithm based on regularized least-squares regression~\cite{sindy, ffx}, which yields compact and physically interpretable
analytical expressions. Unlike previous work on the augmentation of turbulence models~\cite{schmelzer2020discovery, cherroud2022sparse}, the least-squares problem
is not formulated on target values of the corrective terms extracted from the data, but on observable quantities of interest (QoIs), i.e.\ outputs of the RANS
solver, through a CFD-driven training approach~\cite{saidi2022cfd}.
To this end, the corrections are formulated as combinations of a large number of candidate functions of $\boldsymbol{\eta}$:
\begin{equation}\label{eq:corrective_terms}
    R^P = f(\boldsymbol{\eta}) \exp \Big( - \sum_{i \in \mathcal{I}_P} \Gamma_i \phi_i(\boldsymbol{\eta}) \Big), \qquad
    R^E = g(\boldsymbol{\eta}) \exp \Big( - \sum_{j \in \mathcal{I}_E} \Gamma_j \psi_j(\boldsymbol{\eta}) \Big),
\end{equation}
where $\phi_i$ and $\psi_j$ are candidate basis functions, $\mathcal{I}_P$ and $\mathcal{I}_E$ the corresponding index sets, and the $\Gamma_i$ unknown coefficients
identified by sparse regression. The two factors in each of Eqs.~\eqref{eq:corrective_terms} play distinct roles: the prefactors $f$ and $g$ set the amplitude and
sign of the corrections, that is, how strongly the baseline ramp functions are modified, whereas the exponential factors set their spatial support, that is, where the
modification acts. The latter are designed to vanish in fully turbulent regions, where the eddy-viscosity ratio becomes large, so that the baseline formulation, and
hence the underlying $k$--$\omega$ model, is recovered away from the transitional region. This separation facilitates the interpretation of the identified models.

\subsection{Choice of the features}\label{sec:features}
The features are chosen to satisfy four requirements: i) non-dimensionality, ii) Galilean and rotational invariance, iii) informativeness with respect to the
transition process, and iv) locality. The last requirement, shared by most modern transition models, ensures compatibility with parallel and unstructured CFD
solvers. Specifically, we retain four features, all of which are already used, in some form, in existing transition models~\cite{menter2006transition, menter2015one,
srivastava2021generalizable, fang2024data}.

The first feature, which enters the prefactors of the corrections, is the local pressure-gradient parameter of Menter \textit{et al.}~\cite{menter2015one},
\begin{equation}\label{eq:lambda_theta_menter}
    \eta_0 = \lambda_{\theta} = -7.57 \times 10^{-3} \frac{\partial V}{\partial n} \frac{\rho d_w^2}{\mu} + 0.0128,
\end{equation}
where $\partial V / \partial n$ is the wall-normal derivative of the wall-normal velocity component. In a boundary layer, continuity relates this derivative to the
streamwise acceleration of the flow, so that $\lambda_{\theta}$ is a local surrogate of the classical pressure-gradient parameter $(\theta^2/\nu)\, \mathrm{d}U_e /
\mathrm{d}s$, $\theta$ being the momentum thickness, $U_e$ the boundary-layer edge velocity and $s$ the direction tangent to the wall. This feature is introduced because the baseline model does not
include any explicit dependence on the pressure gradient, which strongly affects transition onset. As in~\cite{menter2015one}, $\lambda_{\theta}$ is bounded to
the interval $[-1, 1]$ for numerical robustness.

The three remaining features enter the exponential factors:
\begin{equation}\label{eq:features}
    \eta_1 = R_t = \frac{\mu_t}{\mu}, \qquad
    \eta_2 = \frac{d_w}{d_w + \sqrt{k}/\omega}, \qquad
    \eta_3 = \frac{\rho d_w^2 \Omega}{\mu}.
\end{equation}
The eddy-viscosity ratio $\eta_1$ is small in laminar boundary layers and much larger than unity in turbulent ones, and thus discriminates between the two regimes;
it already enters the baseline model through $T_{\omega}$ and $F_{\mathrm{turb}}$. Moreover, since $\mu_t/\mu = \sqrt{k}\, \ell_t / \nu$, with
$\ell_t = \sqrt{k}/\omega$ the turbulent length scale and $\nu = \mu/\rho$, $\eta_1$ is, in the freestream, a turbulence Reynolds number built on the velocity and
length scales of the turbulence, analogous to the freestream-turbulence Reynolds number introduced by Fransson \& Shahinfar~\cite{fransson2020effect} to account for
the combined effects of the turbulence intensity and of the integral length scale on bypass transition. The second feature, $\eta_2 = (1 + \ell_t/d_w)^{-1}$,
compares the wall distance with the turbulent length scale. The ratio $\ell_t/d_w$ is proportional to the local, Galilean-invariant estimate of the turbulence
intensity used by Menter \textit{et al.}~\cite{menter2015one}, $Tu_L = 100 \sqrt{2k/3} / (\omega d_w)$, so that $\eta_2$ decreases monotonically as the local
turbulence intensity increases and provides information on the freestream turbulence level, which controls bypass transition. Finally, $\eta_3$ is the vorticity
Reynolds number of van Driest \& Blumer~\cite{van_driest}, whose maximum across a zero-pressure-gradient boundary layer is proportional to the momentum-thickness
Reynolds number~\cite{menter2006transition}. It is the basis of the onset criterion of most local transition models, including the baseline model, in which
$Re_v = \eta_3 / 2.188$. The four features can thus be regarded as local counterparts of the parameters that govern bypass transition in classical
correlations~\cite{langtry2009correlation, fransson2020effect}: the momentum-thickness Reynolds number ($\eta_3$), the freestream turbulence intensity ($\eta_2$) and
length scale ($\eta_1$), and the pressure-gradient parameter ($\eta_0$). These features then correlate with transition onset and extent.

The features are not rescaled further, e.g.\ by statistics of the training fields, so as to avoid any dependence on dataset-specific scaling constants. As a
consequence, they span very different ranges: $\eta_2 \in [0, 1]$ and $\lambda_{\theta} \in [-1, 1]$, whereas $\eta_1$ is much larger than unity in turbulent regions
and $\eta_3$ reaches $\mathcal{O}(10^2)$ to $\mathcal{O}(10^3)$ in a transitional boundary layer. These differences in scale are absorbed by the coefficients of the
basis functions introduced below, which must be interpreted accordingly.

\subsection{Modeling ansatz}\label{sec:ansatz}
The prefactors $f$ and $g$ are modeled as third-order polynomials of the pressure-gradient parameter $\eta_0 = \lambda_{\theta}$ only, so that the amplitude and the
sign of the corrections adapt to the local pressure gradient, whereas the exponential factors depend on the remaining features. Following previous work on sparse
symbolic regression for turbulence modeling~\cite{schmelzer2020discovery, cherroud2022sparse}, the basis functions $\phi_i$ and $\psi_j$ are monomials
\begin{equation}\label{eq:monomials}
    \phi_i = \eta_1^{p_1} \eta_2^{p_2}, \qquad
    \psi_j = \eta_1^{q_1} \eta_3^{q_2}, \qquad p_1, p_2, q_1, q_2 \in \mathbb{Z}_{\geq 0}.
\end{equation}
Since $\eta_1$, $\eta_2$ and $\eta_3$ can vanish, negative exponents are excluded, so that the basis functions remain bounded over the whole range of the features.
All coefficients $\Gamma_i$ of the exponential terms are moreover constrained to be non-negative. Since the monomials are non-negative, the exponential terms
remain bounded within $\mathopen{]}0,1]$, and the magnitude of each correction is bounded by that of its polynomial prefactor, i.e.\ $|R^P| \leq |f|$ and
$|R^E| \leq |g|$. Neither correction can therefore grow without bound in the freestream or in fully turbulent regions, which allows the augmented model to be
integrated in the RANS solver without ad hoc limiters.

Guided by a prior local sensitivity analysis of the outputs to the correction terms, and by the need to keep the dimension of the design space reasonable, the
libraries are restricted as follows: the production basis contains the seven monomials of $\eta_1$ and $\eta_2$ of total degree at most three and partial degrees at
most two, and the destruction basis the six mixed monomials $\eta_1^{q_1} \eta_3^{q_2}$, with $q_1, q_2 \geq 1$, of total degree at most four. Together with the
eight polynomial coefficients, the 21 unknown coefficients are gathered in a single vector $\bm{\Gamma} = (\Gamma_1, \dots, \Gamma_{21})$, and the corrective terms
explicitly read
\begin{align}
    R^P &= (\Gamma_1 + \Gamma_2 \lambda_{\theta} + \Gamma_3 \lambda_{\theta}^2 + \Gamma_4 \lambda_{\theta}^3) \nonumber \\
    &\quad \times \exp \left( - \Gamma_5 \eta_1 - \Gamma_6 \eta_2 - \Gamma_7 \eta_1^2 - \Gamma_8 \eta_1 \eta_2 - \Gamma_9 \eta_2^2 - \Gamma_{10} \eta_1^2 \eta_2 - \Gamma_{11} \eta_1 \eta_2^2 \right), \label{eq:rp_explicit}\\
    R^E &= (\Gamma_{12} + \Gamma_{13} \lambda_{\theta} + \Gamma_{14} \lambda_{\theta}^2 + \Gamma_{15} \lambda_{\theta}^3) \nonumber \\
    &\quad \times \exp \left( - \Gamma_{16} \eta_1 \eta_3 - \Gamma_{17} \eta_1 \eta_3^2 - \Gamma_{18} \eta_1^2 \eta_3 - \Gamma_{19} \eta_1^2 \eta_3^2 - \Gamma_{20} \eta_1^3 \eta_3 - \Gamma_{21} \eta_1 \eta_3^3 \right), \label{eq:re_explicit}
\end{align}
so that $\mathcal{I}_P = \{5, \dots, 11\}$ and $\mathcal{I}_E = \{16, \dots, 21\}$ in Eq.~\eqref{eq:corrective_terms}. Such a full expansion is overly complex
and prone to overfitting. It is therefore pruned automatically, the relevant components of $\bm{\Gamma}$ being identified from high-fidelity data through the sparse
symbolic regression procedure detailed in Section~\ref{sec:cfd_driven_identification}.

\section{Datasets}\label{sec:database}
Model training relies on a recent wall-resolved LES database of boundary layers undergoing transition induced by freestream turbulence
(FST)~\cite{zemmour2026high}, whose time-averaged statistics are publicly available~\cite{zemmour_2025_171662106}. The numerical set-up and the validation against
the ERCOFTAC T3 experiments~\cite{roach1992influence, savill1993evaluating} are presented in a companion paper~\cite{zemmour2026high}; only the elements required
for the present study are recalled here.

The simulations were performed with the in-house code \textit{MUSICAA}, which solves the compressible Navier--Stokes equations with high-order centered finite
differences (tenth order for the Eulerian fluxes, fourth order for the viscous fluxes) and a four-stage Runge--Kutta scheme with implicit residual
smoothing~\cite{bienner2024multiblock}. Explicit selective filtering regularizes the unresolved scales, and freestream turbulence is injected with a synthetic
generation technique based on random Fourier modes~\cite{bienner2024influence}, calibrated so as to reproduce the
turbulence intensity at the leading edge of the plate. Due to the use of a compressible flow solver, the simulations were performed at higher Mach numbers than the experiments and the geometry was rescaled to match the experimental Reynolds number. Details of the simulated flow conditions and grid resolutions are given in Table~\ref{tab:inlet_mesh_ercoftac}.
The database includes full fields of mean flow quantities and turbulent statistics, as well as postprocessed wall quantities, including the skin friction coefficient.

Two families of configurations are considered (Table~\ref{tab:inlet_mesh_ercoftac}). The T3A and T3B cases are zero-pressure-gradient (ZPG) boundary layers under
moderate and high freestream turbulence, respectively. The T3C cases combine moderate freestream turbulence with a streamwise pressure gradient, first favorable
and then adverse, imposed by shaping the upper wall of the domain according to the profile of Suluksna \textit{et al.}~\cite{suluksna2009correlations}:
\begin{equation}\label{eq:t3cr_upper_wall}
    \frac{h}{D} = \min \left( 1.231x^6 - 6.705x^5 + 14.061x^4 - 14.113x^3 + 7.109x^2 - 1.900x + 0.950,\ 1 \right),
\end{equation}
where $D$ is the height of the domain at the inlet, $h$ the local height and $x$ the dimensional streamwise coordinate measured from the leading edge.
The training set comprises the T3A, T3B, T3C3 and T3C5 cases: the first two inform the model about the effect of the freestream turbulence intensity, while the
last two correspond to transition under an adverse and a favorable pressure gradient, respectively.

Two held-out cases are used for testing: the T3C2 configuration, whose different pressure-gradient history leads to transition in the adverse-pressure-gradient
region, and an additional ZPG flat-plate case, referred to as ZPG-val, with a freestream turbulence intensity intermediate between those of T3A and T3B but a much
larger integral length scale~\cite{bienner2024influence}, whose statistics are also available on Zenodo~\cite{bienner_2024_12915280}.

\begin{table}[!htbp]
    \centering
    \footnotesize
    \setlength{\tabcolsep}{4pt}
    \renewcommand{\arraystretch}{1.1}
    \begin{tabular}{lcccccc}
        \toprule
                                           & T3A          & T3B          & T3C3         & T3C5         & T3C2         & ZPG-val \\
        \midrule
        Pressure gradient                  & zero         & zero         & variable     & variable     & variable     & zero    \\
        $Tu^{x=0}_{\mathrm{exp}}$ (\%)     & 3            & 6            & 3            & 3            & 3            & --      \\
        $Tu^{x=0}_{x}$ (\%)                & 3.2          & 6.2          & 2.7          & 3.1          & 2.7          & 4.0     \\
        $Re_{L_f}$                         & 2261         & 7914         & 713          & 2319         & 1249         & 14350   \\
        $M_{\mathrm{in}}^{\mathrm{exp}}$   & 0.0157       & 0.0274       & 0.0108       & 0.0245       & 0.0146       & --      \\
        $M_{\mathrm{in}}$                  & 0.5          & 0.5          & 0.2          & 0.2          & 0.2          & 0.1     \\
        Grid points                        & 13.2\,M      & 46\,M        & 44.1\,M      & 186\,M       & 88\,M        & 270\,M  \\
        $\Delta x^+,\ \Delta y^+,\ \Delta z^+$ & 20, 0.9, 12 & 23, 1.0, 13 & 19, 1.5, 11 & 14, 1.4, 14 & 16, 1.5, 10 & 26, 0.9, 10 \\
        Use                                & training     & training     & training     & training     & test         & test    \\
        \bottomrule
    \end{tabular}
    \caption{Flow cases of the LES database. For each case: streamwise pressure gradient; turbulence intensity at the leading edge of the plate in the
    experiments ($Tu^{x=0}_{\mathrm{exp}}$) and streamwise turbulence intensity at the same location in the simulations ($Tu^{x=0}_{x}$); Reynolds number based on the integral length scale of the inflow turbulence,
    $Re_{L_f}$; inlet Mach number in the experiments ($M_{\mathrm{in}}^{\mathrm{exp}}$) and in the simulations ($M_{\mathrm{in}}$); number of grid points and
    near-wall resolution in wall units. Experimental values from~\cite{roach1992influence, savill1993evaluating}, numerical values
    from~\cite{zemmour2026high, bienner2024influence}.}
    \label{tab:inlet_mesh_ercoftac}
\end{table}

\section{CFD-driven symbolic identification}\label{sec:cfd_driven_identification}
The coefficient vector $\bm{\Gamma}$ is identified from the high-fidelity database with a surrogate-assisted CFD-driven approach, inspired by Ben Hassan
Sa\"idi \textit{et al.}~\cite{saidi2022cfd}. Unlike \textit{a priori} methods, which learn the correction from high-fidelity fields under the frozen-field
assumption, this strategy embeds the RANS solver in the optimization loop. Because the identification is driven by the RANS outputs themselves, inconsistencies
between the modeled terms and the resulting flow fields are mitigated, while the surrogate model greatly reduces the computational cost.

\subsection{Objective function and regularization}
Model training minimizes the mean squared error (MSE) between the high-fidelity data $Q_i$ and the RANS predictions $Q_i^{\mathrm{RANS}}(\bm{\Gamma})$, normalized by
the error of the baseline model:
\begin{equation}\label{eq:loss_mse}
    \mathcal{L}_{\mathrm{MSE}}(\bm{\Gamma}) = \frac{1}{M} \sum_{i=1}^{M} \frac{\left\| Q_i - Q^{\mathrm{RANS}}_i(\bm{\Gamma}) \right\|_{2}^2}{\left\| Q_i - Q^{\mathrm{RANS, baseline}}_i \right\|_{2}^2},
\end{equation}
where the index $i$ runs over the $M$ target fields, i.e.\ the QoIs of all the training cases, and $\|\cdot\|_2$ is the discrete $L_2$ norm over the masked region
defined below. The normalization makes the contributions of the different QoIs and cases comparable. To prevent overfitting and to promote interpretability through
sparsity, we use the elastic-net regularization~\cite{zou2005regularization}, which combines the least absolute shrinkage and selection operator
(Lasso)~\cite{tibshirani1996regression} and ridge~\cite{hoerl1970ridge} penalties:
\begin{equation}\label{eq:elastic-net}
    \mathcal{L}(\bm{\Gamma}; \lambda_{\text{reg}}, r) = \mathcal{L}_{\mathrm{MSE}}(\bm{\Gamma}) + \lambda_{\text{reg}} \left( r \|\bm{\Gamma}\|_1 + \frac{1-r}{2} \|\bm{\Gamma}\|_2^2 \right),
\end{equation}
where $\|\bm{\Gamma}\|_1 = \sum_j |\Gamma_j|$ and $\|\bm{\Gamma}\|_2^2 = \sum_j \Gamma_j^2$, the sums running over the $N_p = 21$ coefficients,
$\lambda_{\text{reg}} \geq 0$ controls the overall regularization strength, and $r \in [0, 1]$ the mixing ratio between the $L_1$ and $L_2$ penalties.

An intrinsic difficulty in learning transition models is that transition is spatially localized: it is confined to boundary layers and wakes, and leaves the
freestream unaffected. Most of the field data are therefore uninformative for the transition process, and would severely unbalance the loss function. To exclude the
freestream when computing the losses of Eqs.~\eqref{eq:elastic-net} and~\eqref{eq:elastic-net_bilevel}, the normalized strain-rate magnitude
\begin{equation}\label{eq:norm_strain_rate_magnitude}
    S^* = \frac{\sqrt{S_{ij}S_{ij}}}{\sqrt{S_{ij}S_{ij}} + \sqrt{W_{ij} W_{ij}}}
\end{equation}
is used as a mask that identifies the high-shear regions where transition occurs. In sheared regions such as boundary layers and wakes, the strain-rate and
rotation-rate magnitudes are comparable, so that $S^* \approx 0.5$, whereas in the freestream the rotation rate rapidly drops to zero and $S^*$ tends to unity. Only
points with $S^* < 0.8$ are retained in the loss computation, which excludes the freestream while preserving the boundary-layer and wake regions.

The identified models are assessed a posteriori with the improvement score of Cherroud \textit{et al.}~\cite{cherroud2022sparse}, adapted so that each QoI and each
case carry the same weight. The score for the $i$-th QoI reads
\begin{equation}\label{eq:improv_score_qoi}
    m_{Q_i} = \frac{1}{N_c} \sum_{j=1}^{N_c} \left( 1 - \frac{\left\| Q_{i,j} - Q^{\mathrm{RANS}}_{i,j}(\bm{\Gamma}) \right\|_{2}^2}{\left\| Q_{i,j} - Q^{\mathrm{RANS, baseline}}_{i,j} \right\|_{2}^2} \right),
\end{equation}
where $N_c$ is the number of cases considered and $Q_{i,j}$ the $i$-th QoI of the $j$-th case, and the global score is the average over the $N_{\mathrm{QoI}}$ QoIs,
\begin{equation}\label{eq:global_improv_score}
    m_{\mathrm{model}} = \frac{1}{N_{\mathrm{QoI}}} \sum_{i=1}^{N_{\mathrm{QoI}}} m_{Q_i}.
\end{equation}
A positive score indicates an improvement over the baseline model, a negative score a degradation, and a score of $100\%$ perfect agreement with the high-fidelity
data. In the following, scores are reported for the two training QoIs, the mean streamwise velocity $\overline{u}$ and its wall-normal derivative
$\partial \overline{u}/\partial y$ (Section~\ref{sec:surrogate}), and for the skin-friction coefficient $C_f$, which is not used for training and thus provides an
additional check of the identified models.

\subsection{Surrogate modeling}\label{sec:surrogate}
Learning a model in a CFD-driven setting requires many solver evaluations for different values of the coefficients as the minimization of the loss of
Eq.~\eqref{eq:elastic-net} proceeds; in sparse symbolic regression, this minimization is moreover repeated for each combination of hyperparameters to be tested.
Using the full RANS solver for every evaluation would make training computationally prohibitive, and a surrogate model is therefore introduced. A standard choice
consists in learning the input--output relation between the coefficients and the loss, $\bm{\Gamma} \mapsto \mathcal{L}_{\mathrm{MSE}}(\bm{\Gamma})$. This lumps
the error information into a single scalar and discards its spatial distribution. We instead introduce a field surrogate that reconstructs the full spatial
distributions of the QoIs, an approach shown to be more effective than emulating the loss for high-dimensional problems~\cite{carlozo2025bayesian}.

Most transition models are calibrated against a limited set of experimental skin-friction distributions, which are generally not measured directly but derived from
measured near-wall velocity profiles~\cite{roach1992influence, savill1993evaluating}. Here, the QoIs used for training are the mean streamwise velocity
$\overline{u}$ and its wall-normal derivative $\partial \overline{u} / \partial y$, the latter being computed from the mean velocity fields with high-order finite
differences. The surrogate, referred to as POD--Kriging, combines a proper orthogonal decomposition (POD) of RANS fields obtained for different values of
$\bm{\Gamma}$ with a Kriging (Gaussian-process) regression of the POD coefficients as functions of $\bm{\Gamma}$.

In practice, $\bm{\Gamma}$ is sampled with a space-filling Sobol sequence~\cite{sobol1967distribution} generated with the SMT toolbox~\cite{saves2024smt}. A total of
1024 samples is propagated through the RANS solver for each training case, yielding 1024 parameter-dependent fields per case. For a given case, the sampled fields are
projected onto a POD basis obtained by singular value decomposition (SVD), and approximated as
\begin{equation}\label{eq:pod}
    Q^{\mathrm{RANS}}(\bm{\Gamma}, \bm{x}) \approx Q_0(\bm{x}) + \sum_{k=1}^{K} a_k(\bm{\Gamma}) \Phi_k (\bm{x}),
\end{equation}
where $Q_0$ is the mean field and $\Phi_k$ are the $K$ retained POD modes. Each modal coefficient is then approximated by a Kriging model
$\hat{a}_k: \bm{\Gamma} \mapsto a_k(\bm{\Gamma})$. The construction and accuracy of the surrogates are detailed in Appendix~\ref{sec:appendix_surrogate}.

\subsection{Hyperparameter optimization strategies}
The effectiveness of the elastic-net regularization depends critically on the hyperparameters $\lambda_{\text{reg}}$ and $r$ in Eq.~\eqref{eq:elastic-net}. The standard cross-validated grid search
is outlined below, followed by an alternative approach based on a bi-level formulation of the elastic-net problem.

\subsubsection{Grid search with clustering}\label{eq:standard_cv}
The standard practice is to select $\lambda_{\text{reg}}$ and $r$ by grid search: the hyperparameter space is discretized, a model is trained for each combination,
and the best combination is selected by cross-validation. To account for possible discrepancies between the surrogate and the actual RANS solutions, the
cross-validation is performed by propagating the trained models through the RANS solver and comparing their predictions with the high-fidelity data. Because the
grid points are chosen arbitrarily, however, this discrete search may miss the optimum. Here, the hyperparameter space is discretized with nine values of
$\lambda_{\text{reg}} \in \{10^{-7}, 10^{-6}, \dots, 10 \}$ and eight values of $r \in \{ 0, 0.1, 0.25, 0.4, 0.5, 0.75, 0.9, 1 \}$, yielding 72
combinations.
For each combination, the loss of Eq.~\eqref{eq:elastic-net} is minimized on the four training cases with the differential evolution
algorithm~\cite{price2005differential} of the Python package \texttt{pymoo}~\cite{blank2020pymoo}, with a population of 300 individuals evolved over at most 1500
generations. The search is stopped when the loss falls below $10^{-5}$ or when the best individual has not improved over 50 consecutive generations. Most runs
converge well before the maximum budget, after 400 generations on average; each run uses a single core of a workstation with about 4~GB of memory and takes about
30~min.

The resulting 72 candidate models often share very similar structures and coefficient values. Instead of cross-validating all of them on the four training cases, we
first group them with a $k$-means clustering algorithm~\cite{lloyd1982least}, and only the cluster centroids, which represent distinct model forms, are propagated
through the solver and compared with the high-fidelity data (Appendix~\ref{sec:appendix_model_selection}). The centroid with the best skin-friction accuracy over all training cases, denoted $M_{G.S}$, is retained for the analyses that follow.

\subsubsection{Bi-level optimization}\label{sec:bilevel_optimisation}
To automate the hyperparameter tuning, and to avoid an arbitrary discretization of the hyperparameter space, multiple independent optimizations and a manual
cross-validation, we adopt a bi-level optimization framework~\cite{bonnel2019post}, which explores the hyperparameter space continuously.
Hyperparameter selection is treated as the \textit{leader} (upper-level) problem, which minimizes the validation error, while the
identification of the coefficients is the \textit{follower} (lower-level) problem. To this end, the loss of Eq.~\eqref{eq:elastic-net} is rewritten in terms of two
independent hyperparameters, $\alpha = \lambda_{\text{reg}} r$ and $\beta = \lambda_{\text{reg}} (1-r)/2$, which weight the $L_1$ and $L_2$ penalties, respectively:
\begin{equation}\label{eq:elastic-net_bilevel}
    \tilde{\mathcal{L}}(\bm{\Gamma}; \alpha, \beta) = \mathcal{L}_{\mathrm{MSE}}(\bm{\Gamma}) + \alpha \sum_{j=1}^{N_p} |\Gamma_j| + \beta \sum_{j=1}^{N_p} \Gamma_j^2 .
\end{equation}
The bi-level problem then reads
\begin{equation}\label{eq:bilevel}
    \min_{\alpha, \beta \geq 0} \ \mathcal{L}_{\mathrm{MSE}}^{\mathrm{val}}\big(\bm{\Gamma}^*(\alpha, \beta)\big)
    \quad \text{subject to} \quad
    \bm{\Gamma}^*(\alpha, \beta) = \arg \min_{\bm{\Gamma}} \tilde{\mathcal{L}}^{\mathrm{train}}(\bm{\Gamma}; \alpha, \beta),
\end{equation}
where the superscripts indicate that the losses are evaluated on the validation and training cases, respectively. Both levels are solved by differential
evolution with the surrogate models in the loop.

To promote generalization to unseen flow conditions, the validation set is built with a leave-one-case-out cross-validation (LOCOCV) strategy rather than a random
split. Whereas standard cross-validation partitions individual samples at random, LOCOCV partitions the data at the level of flow cases, so that no observation of a
given case appears in both the training and validation sets. This prevents data leakage and provides a more demanding assessment of the ability of the model to
generalize to unseen flow regimes. The four training cases thus yield four training/validation splits, and the bi-level problem is solved independently for each of
them; the model that performs best on its validation case after propagation through the RANS solver, denoted $M_{B.L}$, is retained
(Appendix~\ref{sec:appendix_lococv}).
%

\section{Results on flat-plate boundary layers}\label{sec:results}
This section compares the models $M_{G.S}$ and $M_{B.L}$ identified with the two hyperparameter optimization strategies, and assesses them against the LES data and
the baseline $k$--$\omega$--$\gamma$ model on the training and test cases of Section~\ref{sec:database}.

\subsection{Identified model structures}\label{sec:identified_models}
\begin{figure}[!htbp]
    \centering
    \includegraphics[width=\linewidth]{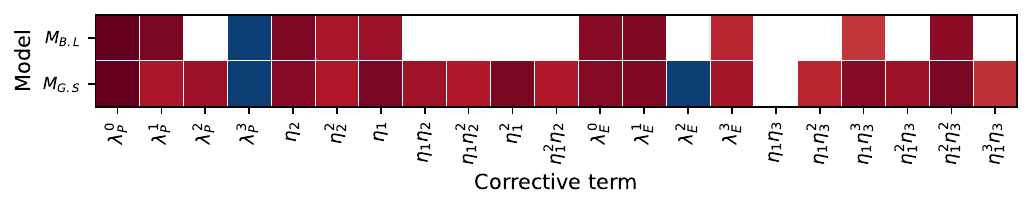}
    \caption{Structure of the models identified with the bi-level ($M_{B.L}$, top row) and grid-search ($M_{G.S}$, bottom row) strategies. Each column
    corresponds to one of the 21 candidate terms of Eqs.~\eqref{eq:rp_explicit}--\eqref{eq:re_explicit}: $\lambda_P^n$ and $\lambda_E^n$ denote the $n$-th order
    terms of the production and destruction prefactors, and the remaining labels the monomials of the production ($\eta_1$, $\eta_2$) and destruction
    ($\eta_1$, $\eta_3$) exponents. White cells correspond to terms pruned by the sparse regression; colored cells to active terms, the color indicating the value of
    the coefficient. The color scale goes from -1 to 1 with a step of 0.1.}
    \label{fig:models_structures}
\end{figure}

The structures of the two models are compared in Fig.~\ref{fig:models_structures}, and their analytical forms are given in Eqs.~\eqref{eq:model_gs}
($M_{G.S}$) and~\eqref{eq:model_bl} ($M_{B.L}$), respectively.
\begin{equation}\label{eq:model_gs}
    \begin{aligned}
        R^P &= \left( 1.704\times 10^{-2} + 2.647\times 10^{-1} \lambda_{\theta} + 4.040\times 10^{-1} \lambda_{\theta}^2 - 9.195\times 10^{-1} \lambda_{\theta}^3 \right) \\
        &\quad \times \exp \left( - \left[ 7.213\times 10^{-1} \eta_2 + 2.592\times 10^{-1} \eta_2^2 + 1.000\, \eta_1 + 3.841\times 10^{-1} \eta_1 \eta_2 \right.\right. \\
        &\qquad\left.\left. + 2.387\times 10^{-1} \eta_1 \eta_2^2 + 9.276\times 10^{-1} \eta_1^2 + 2.340\times 10^{-1} \eta_1^2 \eta_2 \right] \right),\\[2mm]
        R^E &= \left( 7.647\times 10^{-1} + 7.820\times 10^{-1} \lambda_{\theta} - 9.628\times 10^{-1} \lambda_{\theta}^2 + 3.217\times 10^{-1} \lambda_{\theta}^3 \right) \\
        &\quad \times \exp \left( - \left[ 1.598\times 10^{-1} \eta_1 \eta_3^2 + 6.853\times 10^{-1} \eta_1 \eta_3^3 + 3.925\times 10^{-1} \eta_1^2 \eta_3 \right.\right. \\
        &\qquad\left.\left. + 9.911\times 10^{-1} \eta_1^2 \eta_3^2 + 1.156\times 10^{-1} \eta_1^3 \eta_3 \right] \right),
    \end{aligned}
\end{equation}
\begin{equation}\label{eq:model_bl}
    \begin{aligned}
        R^P &= \left( 1.637\times 10^{-2} + 9.281\times 10^{-1} \lambda_{\theta} - 9.478\times 10^{-1} \lambda_{\theta}^3 \right) \\
        &\quad \times \exp \left( - \left[ 8.595\times 10^{-1} \eta_2 + 2.967\times 10^{-1} \eta_2^2 + 4.182\times 10^{-1} \eta_1 \right] \right),\\[2mm]
        R^E &= \left( 6.953\times 10^{-1} + 8.859\times 10^{-1} \lambda_{\theta} + 1.622\times 10^{-1} \lambda_{\theta}^3 \right) \\
        &\quad \times \exp \left( - \left[ 1.058\times 10^{-1} \eta_1 \eta_3^3 + 5.577\times 10^{-1} \eta_1^2 \eta_3^2 \right] \right).
    \end{aligned}
\end{equation}
The grid-search model retains 20 of the 21 candidate coefficients, discarding only the $\eta_1 \eta_3$ monomial of the destruction exponent: at the selected
hyperparameter pair, the elastic-net penalty has practically failed to sparsify the correction. The bi-level model retains 11 coefficients, and its support is a
subset of that of $M_{G.S}$ in both corrections. The bi-level procedure therefore does not identify a different correction, but prunes the one contained in
$M_{G.S}$: the quadratic terms of both prefactors are removed, the production exponent reduces to a separable form, in which the correction is damped independently
by the eddy-viscosity ratio $\eta_1$ and by the ratio of wall distance to turbulent length scale $\eta_2$, and the destruction exponent reduces to the two terms
$\eta_1 \eta_3^3$ and $\eta_1^2 \eta_3^2$. This difference in sparsity results from the selected hyperparameters. The pair retained by the grid search,
$\lambda_{\mathrm{reg}} = 0.01$ and $r = 0.5$, corresponds to $\alpha = 0.005$ and $\beta = 0.0025$ in the notation of Eq.~\eqref{eq:elastic-net_bilevel},
whereas the bi-level procedure selects $\alpha = 0.0782$ and $\beta = 0.3727$ for the retained model, obtained with T3A as validation case. The bi-level
procedure thus selects a much larger $L_1$ penalty, which explains the sparsity of the resulting model.

The retained coefficients confirm that the two models share a common core. The constant terms of the production and destruction prefactors agree to within $4\%$
and $10\%$, respectively, the cubic coefficient of the production prefactor to within $3\%$, the linear coefficient of the destruction prefactor to within $13\%$,
and the coefficients of the $\eta_2$ and $\eta_2^2$ terms of the production exponent to within about $20\%$ and $15\%$. Two procedures based on different
hyperparameter strategies and different training subsets thus converge to the same leading-order correction, which suggests that these directions of the parameter
space are constrained by the data rather than by the tuning strategy. The largest differences concern the two high-order terms of the destruction exponent, whose
coefficients differ by factors of about $6.5$ and $1.8$, and the $\eta_1$ term of the production exponent, whose role in $M_{G.S}$ is shared with the four $\eta_1$
monomials pruned by the bi-level procedure.

\subsection{Application to training and test cases}\label{sec:training_cases}
\begin{figure}[!htbp]
    \centering
    \begin{subfigure}{0.49\textwidth}
        \includegraphics[width=\linewidth]{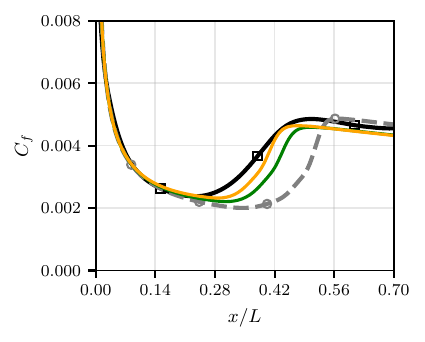}
        \caption{T3A, skin-friction coefficient}
        \label{fig:results_t3a_cf}
    \end{subfigure}\hfill
    \begin{subfigure}{0.49\textwidth}
        \includegraphics[width=\linewidth]{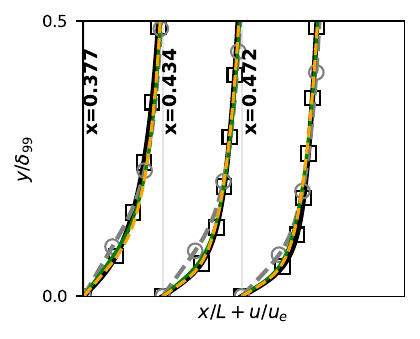}
        \caption{T3A, velocity profiles}
        \label{fig:results_t3a_uu}
    \end{subfigure}\\[2mm]
    \begin{subfigure}{0.49\textwidth}
        \includegraphics[width=\linewidth]{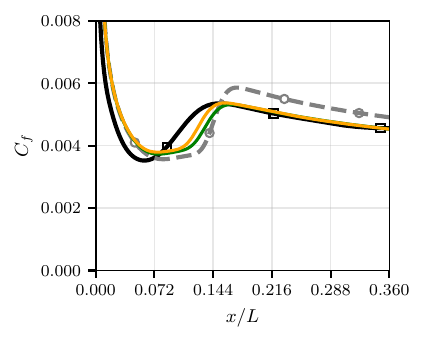}
        \caption{T3B, skin-friction coefficient}
        \label{fig:results_t3b_cf}
    \end{subfigure}\hfill
    \begin{subfigure}{0.49\textwidth}
        \includegraphics[width=\linewidth]{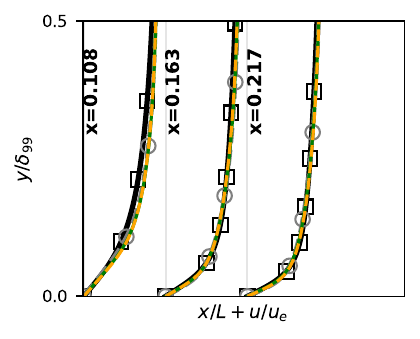}
        \caption{T3B, velocity profiles}
        \label{fig:results_t3b_uu}
    \end{subfigure}
    \caption{Zero-pressure-gradient training cases T3A (top, validation case of $M_{B.L}$) and T3B (bottom): skin-friction coefficient along the plate (left) and mean-velocity profiles
    $u/u_e$ versus $y/\delta_{99}$ at three streamwise stations across the transitional region, each profile being shifted by its abscissa $x/L$ (right). LES: black
    line with squares; baseline $k$--$\omega$--$\gamma$ model: gray dashed line with circles; $M_{B.L}$: orange line; $M_{G.S}$: green line.}
    \label{fig:results_zpg}
\end{figure}

\begin{figure}[!htbp]
    \centering
    \begin{subfigure}{0.49\textwidth}
        \includegraphics[width=\linewidth]{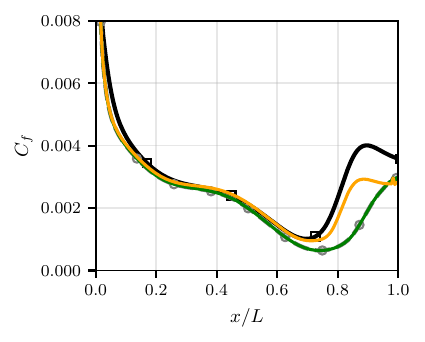}
        \caption{T3C3, skin-friction coefficient}
        \label{fig:results_t3c3_cf}
    \end{subfigure}\hfill
    \begin{subfigure}{0.49\textwidth}
        \includegraphics[width=\linewidth]{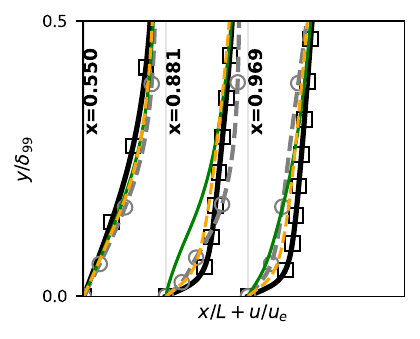}
        \caption{T3C3, velocity profiles}
        \label{fig:results_t3c3_uu}
    \end{subfigure}\\[2mm]
    \begin{subfigure}{0.49\textwidth}
        \includegraphics[width=\linewidth]{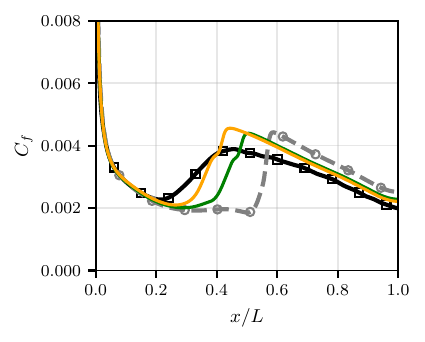}
        \caption{T3C5, skin-friction coefficient}
        \label{fig:results_t3c5_cf}
    \end{subfigure}\hfill
    \begin{subfigure}{0.49\textwidth}
        \includegraphics[width=\linewidth]{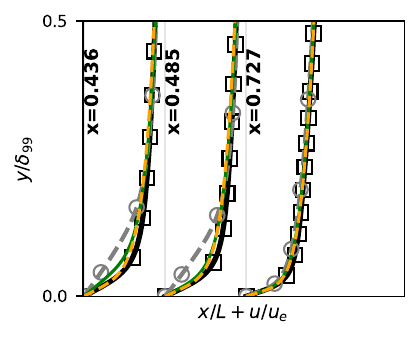}
        \caption{T3C5, velocity profiles}
        \label{fig:results_t3c5_uu}
    \end{subfigure}
    \caption{Pressure-gradient training cases T3C3 (top, transition in the adverse-pressure-gradient region) and T3C5 (bottom,
    transition in the favorable-pressure-gradient region): skin-friction coefficient (left) and mean-velocity profiles (right). Lines as in
    Fig.~\ref{fig:results_zpg}.}
    \label{fig:results_nzpg}
\end{figure}

Table~\ref{tab:improv_scores} summarizes the improvement scores over the four cases used for the identification, T3A having served as validation case for
$M_{B.L}$, and over the two test cases. On the training cases, both models improve on the baseline for all QoIs, $M_{B.L}$ by $28.3\%$ on average against $25.2\%$
for $M_{G.S}$, with about half as many active terms. The skin-friction distributions and velocity profiles of Figs.~\ref{fig:results_zpg}
and~\ref{fig:results_nzpg} show that both corrections move the transition onset upstream, toward the LES, on T3A and on the pressure-gradient cases, $M_{B.L}$ being
the closest to the LES. Two shortcomings remain: the transition length is not fully captured in the pressure-gradient cases, and the improvement on T3B is marginal,
although T3B belongs to the training set of both models. Both are discussed in Section~\ref{sec:interpretation}.

\begin{figure}[!htbp]
    \centering
    \begin{subfigure}{0.49\textwidth}
        \includegraphics[width=\linewidth]{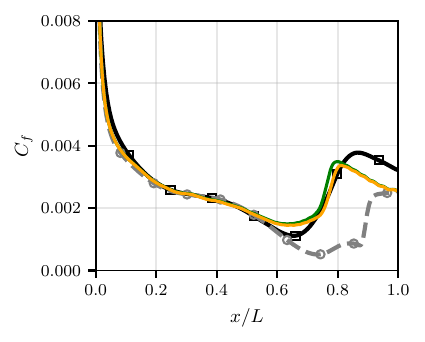}
        \caption{T3C2, skin-friction coefficient}
        \label{fig:results_t3c2_bienner_cf}
    \end{subfigure}\hfill
    \begin{subfigure}{0.49\textwidth}
        \includegraphics[width=\linewidth]{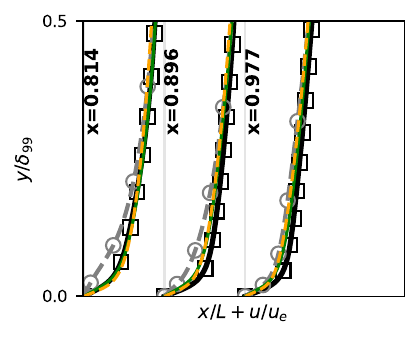}
        \caption{T3C2, velocity profiles}
        \label{fig:results_t3c2_bienner_uu}
    \end{subfigure}\\[2mm]
    \begin{subfigure}{0.49\textwidth}
        \includegraphics[width=\linewidth]{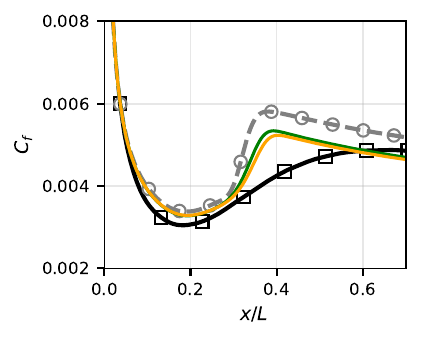}
        \caption{ZPG-val, skin-friction coefficient}
        \label{fig:results_zpg_bienner_cf}
    \end{subfigure}\hfill
    \begin{subfigure}{0.49\textwidth}
        \includegraphics[width=\linewidth]{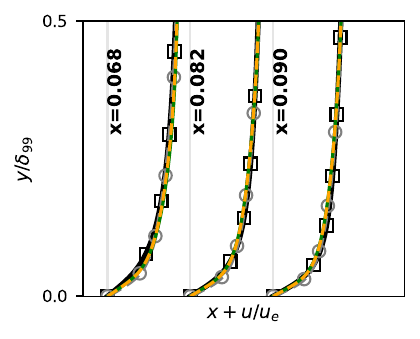}
        \caption{ZPG-val, velocity profiles}
        \label{fig:results_zpg_bienner_uu}
    \end{subfigure}
    \caption{Test cases, not used for training: T3C2 (top, different pressure-gradient history) and ZPG-val (bottom, zero pressure gradient with a large
    integral length scale of the freestream turbulence). Skin-friction coefficient (left) and mean-velocity profiles (right); lines as in
    Fig.~\ref{fig:results_zpg}.}
    \label{fig:results_test}
\end{figure}

Both models also improve on the baseline for the two unseen test cases (Fig.~\ref{fig:results_test}): on T3C2, they move the transition onset upstream, close to the
LES, whereas the baseline transitions too late; on ZPG-val, they reduce the skin-friction overshoot of the baseline, although transition remains more abrupt than in
the LES. The margin between the two models is small, and of the same order as on the training cases. The benefit of the bi-level procedure therefore lies less in a
gain in accuracy than in the automatic selection of a model that is as accurate while being about twice as sparse, which improves its interpretability and, as
discussed in Section~\ref{sec:interpretation}, narrows the range of some of its corrections outside the training conditions.

\begin{table}[!htbp]
    \centering
    \small
    \begin{tabular}{llcccc}
        \toprule
        Cases & Model & $m_{\overline{u}}$ (\%) & $m_{\partial \overline{u} / \partial y}$ (\%) & $m_{C_f}$ (\%) & $m_{\mathrm{model}}$ (\%) \\
        \midrule
        Training & $M_{G.S}$ & 23.83 & 30.46 & 21.26 & 25.18 \\
                                        & $M_{B.L}$ & 25.15 & 34.42 & 25.22 & 28.26 \\
        \midrule
        Test     & $M_{G.S}$ & 16.46 & 40.70 & 13.01 & 23.39 \\
                                        & $M_{B.L}$ & 18.40 & 41.03 & 14.90 & 24.78 \\
        \bottomrule
    \end{tabular}
    \caption{Improvement scores of the grid-search ($M_{G.S}$) and bi-level ($M_{B.L}$) models with respect to the baseline $k$--$\omega$--$\gamma$ model
    (Eqs.~\eqref{eq:improv_score_qoi}--\eqref{eq:global_improv_score}), averaged over the training cases (T3A, T3B, T3C3, T3C5) and test cases (T3C2, ZPG-val), for the mean streamwise velocity, its wall-normal
    derivative and the skin-friction coefficient; $m_{\mathrm{model}}$ is the mean of the three scores. Positive values denote an improvement over the baseline, and
    $100\%$ a perfect agreement with the LES. For $M_{B.L}$, T3A is the validation case of the leave-one-case-out procedure.}
    \label{tab:improv_scores}
\end{table}

\subsection{Interpretation of the identified corrections}\label{sec:interpretation}
The relative magnitudes of the identified terms clarify the nature of the correction. In both models, the constant term of the destruction prefactor is 42 to 45
times larger than that of the production prefactor: the production correction is a small perturbation of the ramp function $F_{\gamma}$, whereas the destruction
correction substantially modifies $G_{\gamma}$. This asymmetry reflects the roles of the two terms in the transport equation. The production correction acts through
$\rho \Omega (\gamma_{\mathrm{max}} - \gamma) \sqrt{\gamma}$, a factor that is largest precisely where the intermittency starts to grow, so that a small change of the
activation function suffices to shift the onset location; the destruction correction acts through $F_{\mathrm{turb}} \rho \Omega \gamma^{1.5}$ and remains inactive
until the intermittency has grown appreciably.

The two procedures also agree on the response of the corrections to $\lambda_{\theta}$. Both destruction prefactors increase with $\lambda_{\theta}$ with similar
slopes, so that the destruction correction is reinforced in accelerated regions and weakened in decelerated ones. Both production prefactors change sign for
moderately negative values of $\lambda_{\theta}$ (below $\lambda_{\theta} \approx -0.02$ for $M_{B.L}$, and within $-0.31 < \lambda_{\theta} < -0.07$ for $M_{G.S}$),
where the production correction delays rather than promotes the growth of intermittency. This dependence should nevertheless be interpreted with care: since
$\lambda_{\theta}$ is evaluated locally from the wall-normal derivative of the wall-normal velocity, it responds to the thickening of the boundary layer through the
transitional region as well as to the imposed pressure gradient, and the sign reversal cannot be attributed to the latter alone.

The same reading explains the two shortcomings noted above. The transition length is controlled almost entirely by the destruction correction, whose exponent
reduces in $M_{B.L}$ to a combination of $\eta_1 \eta_3^3$ and $\eta_1^2 \eta_3^2$. Since the vorticity Reynolds number $\eta_3$ reaches $\mathcal{O}(10^2)$ to
$\mathcal{O}(10^3)$ in a transitional boundary layer and enters at the cubic power, the exponential drops from one to zero as soon as the eddy-viscosity ratio
exceeds about $10^{-8}$ to $10^{-5}$, depending on the local value of $\eta_3$, so that $R^E$ behaves in practice as an indicator function rather than as a smooth
decay. Such a switch can displace the end of the transitional region but has little authority over its extent, which is consistent with the residual discrepancy in
transition length observed in the pressure-gradient cases. The marginal improvement on T3B points to a limitation of the dictionary rather than of the
hyperparameter tuning: the selected features cannot distinguish the high freestream turbulence of T3B ($Tu = 6\%$) from the moderate levels of the other cases.

Finally, since $\lambda_{\theta}$ enters only through the polynomial prefactors, and since the exponential factors are bounded within $\mathopen{]}0,1]$, the behavior of each
correction outside the training conditions can be inspected directly from its prefactor over the admissible range $\lambda_{\theta} \in [-1, 1]$. The destruction
prefactor of $M_{B.L}$ remains positive down to $\lambda_{\theta} \approx -0.72$, whereas that of $M_{G.S}$ changes sign at $\lambda_{\theta} \approx -0.55$ and would
therefore, under strong adverse pressure gradients, subtract from $G_{\gamma}$ rather than add to it. The production prefactor of $M_{B.L}$ is bounded by about
$0.37$ in magnitude, about three times less than that of $M_{G.S}$.

In summary, the bi-level formulation automatically selects a model that is about twice as sparse as the grid-search model and at least as accurate on both training
and test cases. In the next section, $M_{B.L}$ is applied to the LS89 turbine cascade, where the corrections are evaluated well outside the range of $\lambda_{\theta}$
covered by the T3 cases, in particular under the strong acceleration of the suction side.

\section{Application to a turbine cascade}\label{sec:ls89}
In this section, $M_{B.L}$ is applied to a more complex configuration, the highly loaded LS89 linear turbine cascade studied experimentally at the von Karman
Institute (VKI) by Arts \textit{et al.}~\cite{arts1990aero, arts1992aero}, which is widely used as a benchmark for transition and heat-transfer predictions in
turbomachinery~\cite{benyahia2011transition, pecnik2013assessment, dupuy2020analysis, fornasari2025rans}.

The simulations are performed on a structured nine-block grid extending from $x/c = -1.0$ at the inlet to $x/c = 2.0$ at the outlet, with a first-cell height below
$y^+ = 0.5$ to resolve the transitional boundary layers. No-slip isothermal conditions are applied on the blade surface and periodic conditions on the upper and
lower boundaries of the domain. Riemann-invariant conditions are imposed at the inlet and the static pressure is prescribed at the outlet, so as to reproduce the
experimental conditions.

Four operating points are considered, referred to as MUR129, MUR235, MUR116 and MUR241 in~\cite{arts1990aero} (Table~\ref{tab:ls89_conditions}). They form two pairs:
MUR129 and MUR116 correspond to a low inlet turbulence intensity ($Tu = 0.8\%$, a level at which freestream-turbulence effects are small), MUR235 and MUR241 to a high
one ($Tu = 6\%$). Within each pair, the second case has an outlet Reynolds number about twice as high and a slightly supersonic outlet Mach number
($M_2 \approx 1.09$), with shocks impinging on the rear part of the suction side. The inlet turbulence intensity is set to the experimental value. The inlet specific
dissipation rate, which is not documented in the experiments, is imposed through the eddy-viscosity ratio, with values comparable with those used by Smirnov \&
Smirnovsky~\cite{smirnov2009turbine} in their simulations of the LS89 cascade.

\begin{figure}[!htbp]
    \centering
    \begin{subfigure}{0.49\textwidth}
        \includegraphics[width=\textwidth]{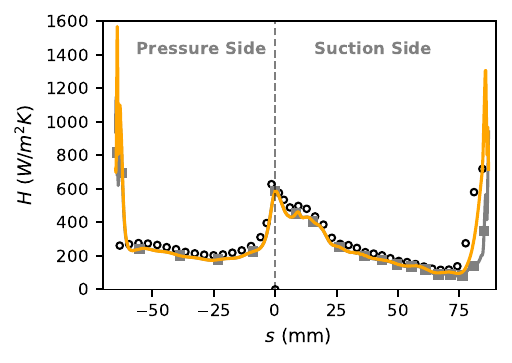}
        \caption{MUR129 ($Tu = 0.8\%$, $M_2 = 0.84$)}
    \end{subfigure}\hfill
    \begin{subfigure}{0.49\textwidth}
        \includegraphics[width=\textwidth]{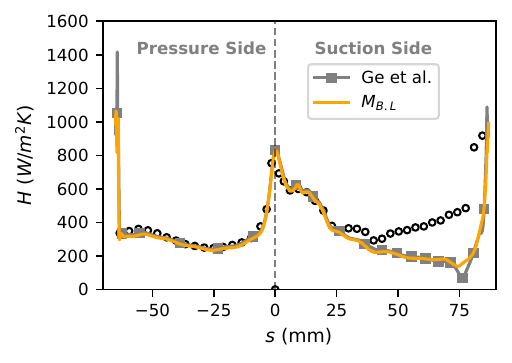}
        \caption{MUR116 ($Tu = 0.8\%$, $M_2 = 1.09$)}
    \end{subfigure}\\[2mm]
    \begin{subfigure}{0.49\textwidth}
        \includegraphics[width=\textwidth]{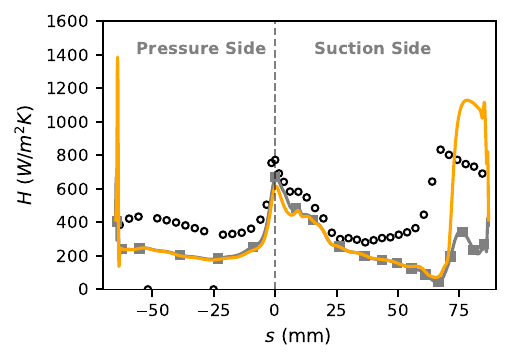}
        \caption{MUR235 ($Tu = 6\%$, $M_2 = 0.93$)}
    \end{subfigure}\hfill
    \begin{subfigure}{0.49\textwidth}
        \includegraphics[width=\textwidth]{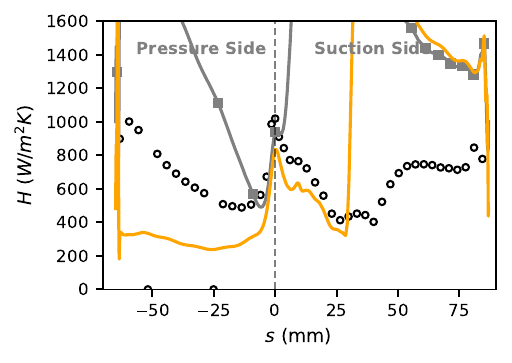}
        \caption{MUR241 ($Tu = 6\%$, $M_2 = 1.09$)}
    \end{subfigure}
    \caption{Heat transfer coefficient $H$ (Eq.~\ref{eq:heat_transfer_coefficient}) along the LS89 blade surface, as a function of the curvilinear
    abscissa $s$ measured from the leading edge ($s < 0$: pressure side; $s > 0$: suction side), for the four operating points. Symbols: experiments of Arts
    \textit{et al.}~\cite{arts1990aero}; gray line with squares: baseline model of Ge \textit{et al.}; orange line: augmented model $M_{B.L}$. Transition appears as a
    sharp rise of $H$ on the suction side.}
    \label{fig:heat_flux_mur}
\end{figure}

The augmented model $M_{B.L}$ is compared with the baseline model of Ge \textit{et al.}~\cite{ge2014bypass}. The predictions are assessed on the distribution along
the blade surface of the heat transfer coefficient
\begin{equation}\label{eq:heat_transfer_coefficient}
    H = \frac{q_w}{T_{0} - T_w},
\end{equation}
where $q_w$ is the wall heat flux, $T_0$ the inlet total temperature and $T_w$ the wall temperature, as in~\cite{arts1990aero}. The distributions of $H$ are shown in
Fig.~\ref{fig:heat_flux_mur}, and the corresponding improvement scores $m_H$, computed as in Eq.~\eqref{eq:improv_score_qoi} with the baseline model as reference, are
reported in Table~\ref{tab:ls89_conditions}.
\begin{table}[!htbp]
    \centering
    \small
    \begin{tabular}{lcccc}
        \toprule
                                                & MUR129 & MUR116 & MUR235 & MUR241 \\
        \midrule
        $Tu$ (\%)                               & 0.8    & 0.8    & 6      & 6      \\
        $M_{\mathrm{in}}$                       & 0.15   & 0.157  & 0.15   & 0.15   \\
        $M_2$                                   & 0.84   & 1.09   & 0.927  & 1.089  \\
        $T_0$ (K)                               & 409.2  & 418.9  & 413.3  & 416.4  \\
        $T_w$ (K)                               & 297.75 & 297.55 & 301.15 & 299.75 \\
        Inlet $\mu_t/\mu$                       & 1.56   & 3      & 15.47  & 20     \\
        \midrule
        $m_H$ of $M_{B.L}$ (\%)                 & 15.2   & 2.4    & 5.1    & 70.1   \\
        \bottomrule
    \end{tabular}
    \caption{LS89 operating points: inlet turbulence intensity, inlet and isentropic outlet Mach numbers, inlet total temperature, wall temperature and
    eddy-viscosity ratio imposed at the inlet of the RANS simulations. Last row: improvement score of $M_{B.L}$ on the heat transfer coefficient $H$, relative to the
    baseline model of Ge \textit{et al.}~\cite{ge2014bypass}.}
    \label{tab:ls89_conditions}
\end{table}

\begin{figure}[!htbp]
    \centering
    \begin{subfigure}{0.49\textwidth}
        \includegraphics[width=\textwidth]{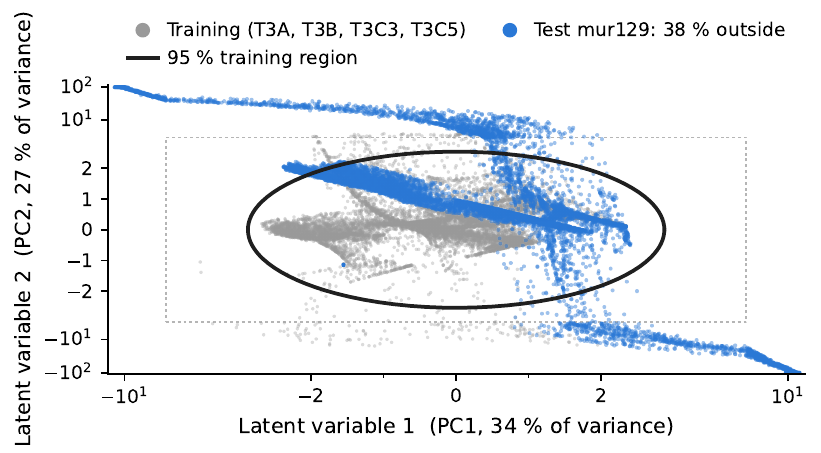}
        \caption{MUR129}
    \end{subfigure}\hfill
    \begin{subfigure}{0.49\textwidth}
        \includegraphics[width=\textwidth]{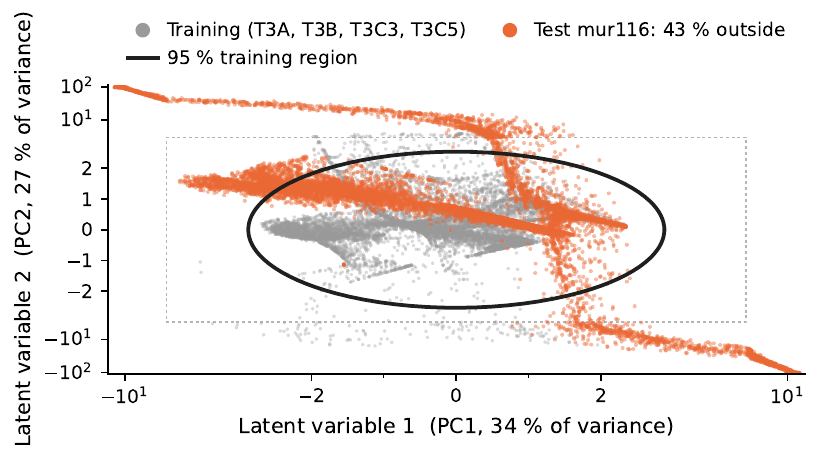}
        \caption{MUR116}
    \end{subfigure}\\[2mm]
    \begin{subfigure}{0.49\textwidth}
        \includegraphics[width=\textwidth]{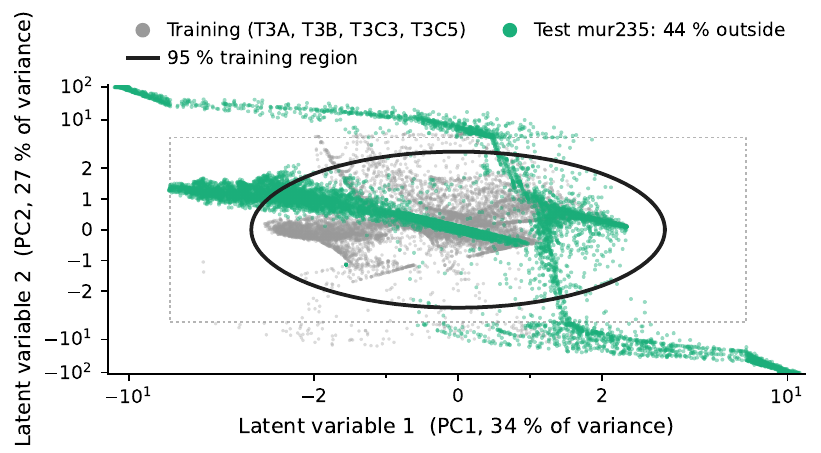}
        \caption{MUR235}
    \end{subfigure}\hfill
    \begin{subfigure}{0.49\textwidth}
        \includegraphics[width=\textwidth]{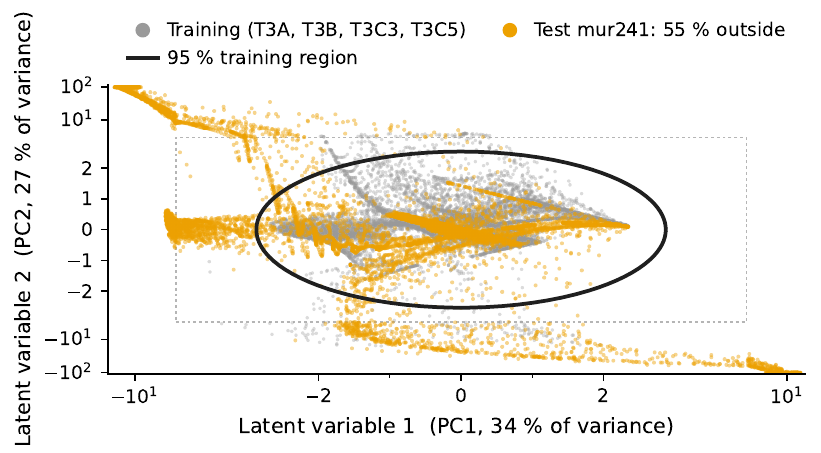}
        \caption{MUR241}
    \end{subfigure}
    \caption{Extrapolation range of the LS89 simulations in feature space: projection of the input features of the corrections onto the first two principal
    components of the training data (gray: T3A, T3B, T3C3 and T3C5; colors: LS89 cases). The ellipse encloses $95\%$ of the training points, and the fraction of LS89
    points lying outside it is indicated in each legend. The dashed rectangle delimits the domain of the training data. Symmetric logarithmic axes.}
    \label{fig:extrapolation}
\end{figure}

This configuration lies well outside the training envelope: compared with the T3 cases, it combines a curved geometry, a much stronger acceleration, higher Reynolds
numbers and, for two of the cases, a transonic regime. The extent of this extrapolation can be quantified by projecting the input features of the corrections onto
a two-dimensional space, following Oulghelou \textit{et al.}~\cite{oulghelou2025blending}. The four features of Section~\ref{sec:features} are used, namely the
pressure-gradient parameter $\lambda_{\theta}$, the eddy-viscosity ratio $\eta_1$, the wall-distance ratio $\eta_2$ and the vorticity Reynolds number $\eta_3$, the
logarithms of $\eta_1$ and $\eta_3$ being taken since they vary over several orders of magnitude. The training points are taken from the T3A, T3B, T3C3 and T3C5
cases, restricted to the masked region ($S^* < 0.8$) used in the training loss, and the test points from the four computational blocks surrounding the blade in
each LS89 case. A principal component analysis (PCA) of the training data provides the two directions along which the training features vary the most, and both
the training and the test points are projected onto these directions (Fig.~\ref{fig:extrapolation}). The region occupied by the training data is represented by the
ellipse that contains $95\%$ of the training points under a Gaussian assumption, its extent along each direction being set by the spread of the training data in that
direction, so that test points falling outside this ellipse correspond to flow conditions not represented in the training set. Between $38\%$ and $55\%$ of the
points of the four LS89 cases lie outside this region, the largest fraction being found for MUR241, which combines high freestream turbulence with a supersonic
outlet. Moreover, the bypass transition process itself is affected by such strong pressure gradients and curvature~\cite{zhao2020bypass}.

For the low-turbulence cases, the baseline and augmented models give nearly identical predictions on the pressure side and on the front part of the suction side,
where the boundary layer remains laminar, and both agree well with the measurements. The differences are confined to the rear part of the suction side. On MUR129,
the baseline model predicts transition only at $s \approx 84$~mm, whereas $M_{B.L}$ triggers it at $s \approx 76$~mm, in closer agreement with the rise of the
measured heat transfer at $s \approx 77$~mm; this earlier onset accounts for the improvement of about $15\%$ over the baseline. On MUR116, the
measurements indicate a gradual rise of $H$ from $s \approx 40$~mm to the trailing edge, characteristic of an extended transitional region. The augmented model hardly
differs from the baseline there: both keep the boundary layer laminar almost up to the trailing edge and underpredict the heat transfer over the rear half of the
suction side, hence an improvement of only $2.4\%$. Neither model captures the transitional region, consistently with the limitation identified in
Section~\ref{sec:interpretation}: the destruction correction behaves as a switch that can move the end of transition but has little control over its length.

The high-turbulence cases are more demanding. On MUR235, the measured transition onset lies at $s \approx 60$--$64$~mm. The baseline model remains essentially
laminar along the suction side, with only a weak rise of $H$ near $s \approx 75$~mm, whereas $M_{B.L}$ predicts transition at $s \approx 68$~mm, closer to the
measurements, but overshoots the measured turbulent level; since the benefit of the earlier onset is offset by this overshoot, the improvement is limited to $5.1\%$.
On MUR241, the baseline model predicts an almost fully turbulent boundary layer on both sides of the blade and overpredicts the heat transfer by a factor of two to
three. The augmented model removes this premature transition: it restores a laminar boundary layer on the pressure side and on the front of the suction side, and
delays the onset to $s \approx 30$~mm, the measurements showing a local minimum of $H$ at $s \approx 40$~mm followed by a sharp rise. The onset is still premature,
but the error is substantially reduced, which explains the improvement of $70\%$ over the baseline. This behavior is consistent with the structure of the identified
correction. In the laminar region, where $\eta_1 \to 0$, the exponential factor of $R^E$ tends to unity, and its prefactor increases with $\lambda_{\theta}$. The
destruction of intermittency is therefore reinforced precisely in the strongly accelerated boundary layer of the suction side, which delays the growth of $\gamma$
and counteracts the premature transition of the baseline model on MUR241. These gains should nevertheless be interpreted with caution: the improvement scores
integrate the whole blade surface, and the large score obtained on MUR241 is dominated by the gross overprediction of the baseline model, whereas the predicted onset
remains upstream of the measured one for this case and the transitional region is not reproduced on MUR116. Finally, neither model reproduces the increase of
the heat transfer induced by the freestream turbulence on the pressure side, which is clearly visible in the measurements of the high-turbulence cases: where the
boundary layer is predicted laminar, the computed heat transfer remains close to that obtained at low turbulence intensity.

Overall, the augmented model, trained exclusively on flat-plate boundary layers, remains numerically robust on a transonic turbine cascade without any additional
limiter, and improves on the baseline in all four cases, most markedly on MUR241, where the baseline transitions prematurely. The remaining discrepancies in the
location and extent of transition on the suction side indicate, however, that the training database should be extended to flows with strong acceleration, surface
curvature and shock--boundary-layer interactions to achieve accurate predictions in turbomachinery configurations.

\section{Conclusions}\label{sec:conclusions}
A CFD-driven sparse symbolic regression framework has been proposed to correct the structural deficiencies of the activation functions of the local,
Galilean-invariant $k$--$\omega$--$\gamma$ transition model of Ge \textit{et al.}~\cite{ge2014bypass}. Each correction is expressed as a polynomial of a local
pressure-gradient parameter multiplied by a bounded exponential of local features. This yields closed-form corrections whose amplitude and spatial support can be
interpreted separately, and an augmented model that runs without ad hoc limiters, including on a transonic turbine cascade.

The CFD-driven procedure, accelerated by POD--Kriging surrogates of the RANS solver, identifies the coefficients from observable quantities, namely the mean velocity
and its wall-normal derivative, without need for an adjoint solver and without target fields for the intermittency. The learned models improve on the baseline model on both training and
unseen flat-plate cases. A bi-level formulation of the elastic-net problem, combined with a leave-one-case-out cross-validation, identifies the regularization
hyperparameters automatically, yields a sparser model than the conventional grid search, with equal or slightly
better accuracy, and removes the need for an arbitrary discretization of the hyperparameter space and for the cross-validation of many candidate models. Both
procedures converge to the same leading-order correction, which consists of a small perturbation of the production term and a substantial, pressure-gradient-dependent
reinforcement of the destruction term.

Applied without retraining to the LS89 cascade, the augmented model improves the heat-transfer predictions of the baseline at all four operating points, most
markedly where the baseline predicts premature transition under high freestream turbulence. The transition length and, at high turbulence intensity, the onset
location remain inaccurately predicted, consistently with the switch-like behavior of the destruction correction and with the large fraction of LS89 flow states
that lie outside the training envelope.

These results suggest several perspectives. First, the training data should be extended to flows with surface curvature, stronger pressure gradients and
shock--boundary-layer interactions, possibly including turbomachinery cases~\cite{zhao2020bypass, dupuy2020analysis}. Second, the dictionary could be enriched with
features able to better discriminate between freestream turbulence levels, and with functional forms giving authority over the transition length. Of note, the
proposed methodology is not specific to the present baseline, and could be applied to other local transition models, such as the one-equation model of Menter
\textit{et al.}~\cite{menter2015one} or the reformulated laminar-kinetic-energy model of Pacciani \textit{et al.}~\cite{pacciani2025reformulation}. 

\backmatter

\bmhead{Acknowledgments}
PC acknowledges funding by the French National Research Agency (ANR) under the France 2030
program, within the framework of the IA Cluster initiative (grant ANR-23-IACL-0007). Access to the HPC resources has been granted by
CINES and TGCC under the allocations A0182A13457 and A0192A01736 made by GENCI (Grand Equipement National de Calcul Intensif).

\section*{Declarations}
\begin{itemize}
    \item \textbf{Funding:} The present research was partly funded by the French National Research Agency (ANR) under the France 2030
program, within the framework of the IA Cluster initiative (grant ANR-23-IACL-0007).
    \item \textbf{Competing interests:} The authors declare no competing interests.
    \item \textbf{Data availability:} The time-averaged statistics of the LES database are available on Zenodo~\cite{zemmour_2025_171662106}; the ZPG-val data are
    available in~\cite{bienner_2024_12915280}.
    \item \textbf{Code availability:} The identification framework is available at \url{https://github.com/LouenasZm/AugTransitionRANS}.
    \item \textbf{Author contributions:} \textit{L. Zemmour}: Methodology, Formal Analysis, Algorithm implementation, Writing – Original Draft. \textit{X. Gloerfelt}: Supervision, Writing – Review \& Editing. \textit{P. Cinnella}: Conceptualization, Supervision, Writing – Review \& Editing
\end{itemize}

\bibliography{sn-bibliography}

\clearpage
\begin{appendices}
%
\section{Construction and accuracy of the surrogate models}\label{sec:appendix_surrogate}
For each training case and QoI, the 1024 samples are split into training, validation and test sets (60/20/20\%), and POD modes are retained up to $99\%$ of the
captured energy. A grid search is then performed over the trend functions (constant, linear, quadratic) and correlation models of the Kriging regression (standard
Kriging and KPLSK, i.e.\ Kriging with a partial-least-squares reduction of the input dimension~\cite{bouhlel2016improving}, both available in
SMT~\cite{saves2024smt}), and the configuration minimizing the error on the validation set is retained. The accuracy of the surrogates is quantified by the
coefficient of determination $R^2$, which measures the quality of the fit on the training samples, and by $Q^2$, which measures the predictive quality on the unseen
test samples:
\begin{equation}
    R^2 = 1 - \frac{\sum_{i=1}^{N} |y_{i,\mathrm{train}} - \hat{y}_{i,\mathrm{train}}|^2}{\sum_{i=1}^{N} |y_{i,\mathrm{train}} - \bar{y}_{\mathrm{train}}|^2},
    \qquad
    Q^2 = 1 - \frac{\sum_{i=1}^{N} |y_{i,\mathrm{test}} - \hat{y}_{i,\mathrm{test}}|^2}{\sum_{i=1}^{N} |y_{i,\mathrm{test}} - \bar{y}_{\mathrm{test}}|^2},
\end{equation}
where $y_i$ is the RANS value, $\hat{y}_i$ the surrogate prediction and $\bar{y}$ the mean of the RANS values over the corresponding set; values close to unity
indicate a good fit and a good predictive capability, respectively. As reported in Table~\ref{tab:surrogate_scores}, both scores exceed $99.8\%$ for all cases and
QoIs.
\begin{table}[!htbp]
    \centering
    \small
    \begin{tabular}{llccc}
        \toprule
        Case & QoI & Kriging model & $R^2$ (\%) & $Q^2$ (\%) \\
        \midrule
        T3A  & $\overline{u}$                       & Kriging & 99.9953 & 99.9856 \\
             & $\partial \overline{u} / \partial y$ & Kriging & 99.9914 & 99.9632 \\
        T3B  & $\overline{u}$                       & KPLSK   & 99.9978 & 99.9887 \\
             & $\partial \overline{u} / \partial y$ & KPLSK   & 99.9981 & 99.986  \\
        T3C3 & $\overline{u}$                       & Kriging & 99.9685 & 99.3043 \\
             & $\partial \overline{u} / \partial y$ & Kriging & 99.966  & 99.7159 \\
        T3C5 & $\overline{u}$                       & Kriging & 99.9952 & 99.9615 \\
             & $\partial \overline{u} / \partial y$ & Kriging & 99.977  & 99.5766 \\
        \bottomrule
    \end{tabular}
    \caption{Accuracy of the POD--Kriging surrogates for each training case and QoI (mean streamwise velocity $\overline{u}$ and its wall-normal derivative
    $\partial \overline{u} / \partial y$): selected Kriging model, coefficient of determination on the training ($R^2$) and on the test samples ($Q^2$).}
    \label{tab:surrogate_scores}
\end{table}
%
\section{Grid search and model selection}\label{sec:appendix_model_selection}
\begin{figure}[!htbp]
    \centering
    \includegraphics[width=0.75\linewidth]{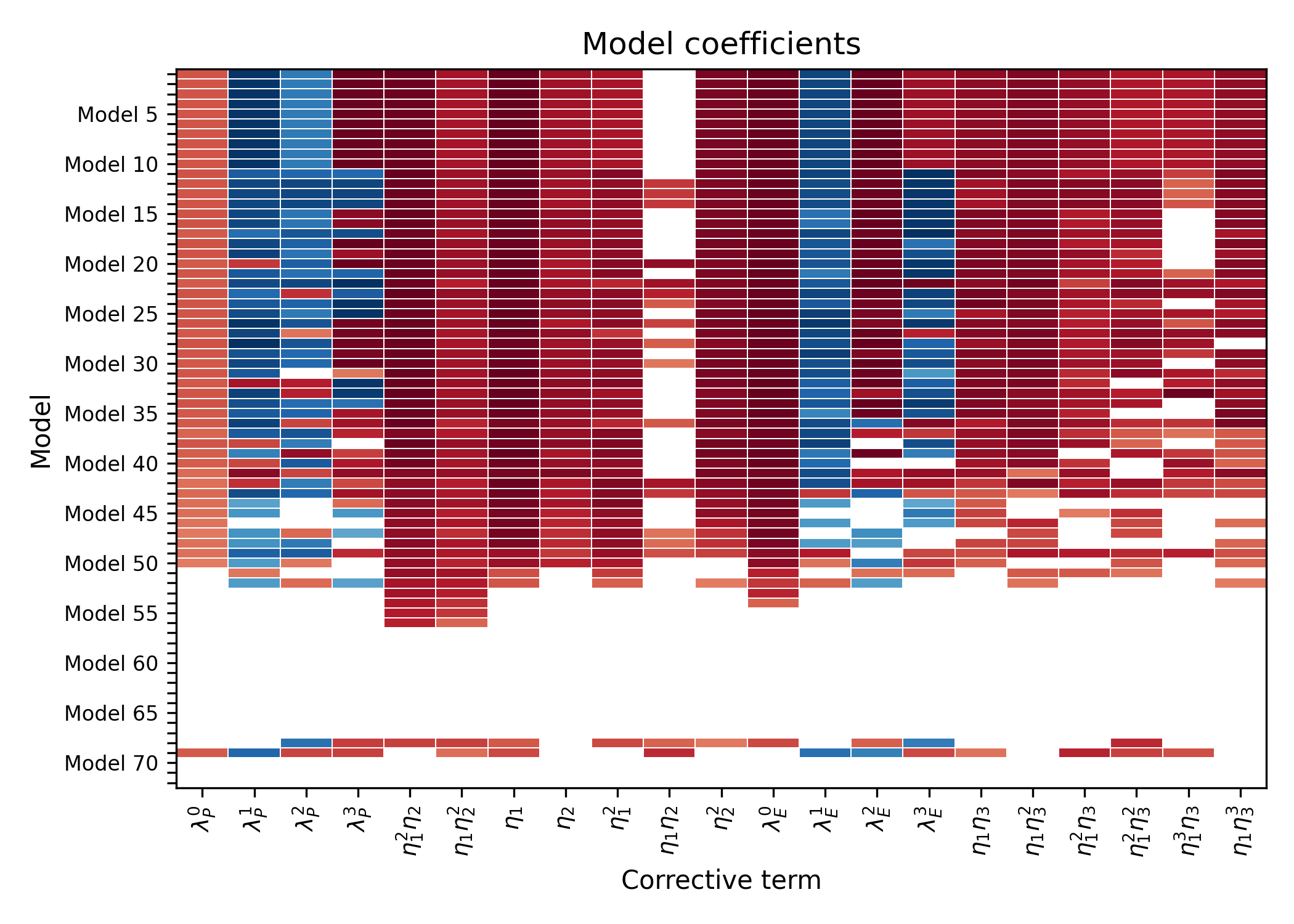}
    \caption{Coefficients of the 72 candidate models obtained by the elastic-net grid search on the ZPG training cases T3A and T3B (one row per hyperparameter
    pair; one column per candidate term, with the notation of Fig.~\ref{fig:models_structures}). Red and blue shades correspond to positive and negative
    coefficients, respectively; white cells correspond to inactive terms.}
    \label{fig:app_elastic_net_all_models}
\end{figure}

This appendix details the grid-search procedure outlined in Section~\ref{eq:standard_cv}. For each of the 72 hyperparameter pairs $(\lambda_{\text{reg}}, r)$ of
the grid, a candidate model is trained by CFD-driven optimization (Fig.~\ref{fig:app_elastic_net_all_models}).

Propagating all the candidates through the RANS solver would be computationally prohibitive. The candidates are therefore first grouped by similarity, and only
one representative model per group is propagated through the solver. Each candidate is represented by its coefficient vector
$\bm{\Gamma} \in \mathbb{R}^{N_p}$, with $N_p = 21$. Rather than clustering the raw coefficient values, each vector is expanded into three complementary descriptors:
(i) a binary mask of the active (non-zero) coefficients, which captures the structure of the model; (ii) the coefficient vector normalized to unit $L_2$ norm, which
captures the relative weights of the active coefficients independently of their overall scale; and (iii) the absolute coefficient values normalized to sum to one,
which capture the relative importance of each term. The three descriptors are concatenated into a single feature vector per model and standardized to zero mean and
unit variance across the 72 models, so that neither a descriptor nor an individual coefficient dominates the distance metric merely because of its scale.

The standardized feature vectors are then partitioned with the $k$-means algorithm~\cite{lloyd1982least}, which assigns each vector to one of $k$ clusters so as to
minimize the total within-cluster sum of squared distances to the cluster centroids. The number of clusters is chosen by evaluating, over a range of candidate
values, the silhouette score~\cite{rousseeuw1987silhouettes}, which measures how well separated and internally coherent the clusters are, together with the
conventional elbow criterion on the $k$-means inertia. The cluster centroids are then propagated through the RANS solver. Since the candidate models are trained on
the velocity field and its wall-normal derivative, and none of the training targets involves the skin-friction coefficient directly, each propagated centroid is
evaluated on the skin-friction coefficient of the full RANS solution, through its MSE with respect to the high-fidelity data. This provides an independent check of
predictive accuracy on a quantity that played no role in the fitting. The global improvement score of Eq.~\eqref{eq:global_improv_score}, which also accounts for
the training QoIs, is used as a check, and both criteria yield the same ranking.

\begin{figure}[!htbp]
    \centering
    \includegraphics[width=0.55\linewidth]{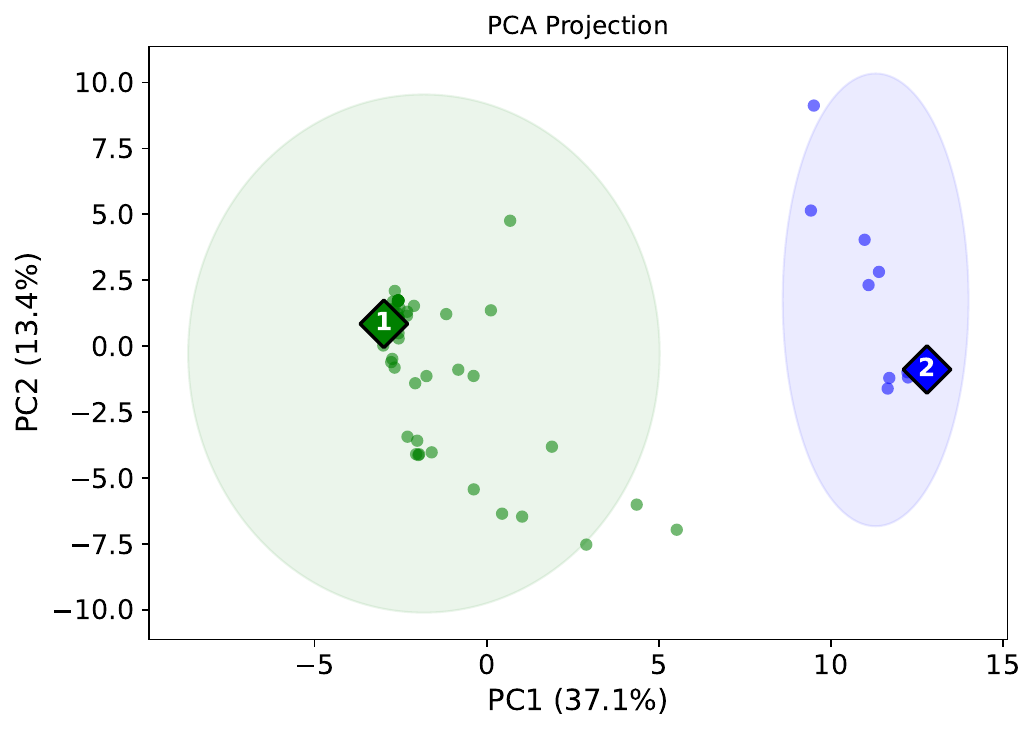}
    \caption{Projection of the standardized descriptors of the 72 candidate models of Fig.~\ref{fig:app_elastic_net_all_models} onto their first two principal
    components. Dots: candidate models, colored by $k$-means cluster; diamonds: the two cluster centroids, $M_1^{\mathrm{ZPG}}$ (1) and $M_2^{\mathrm{ZPG}}$ (2),
    propagated through the RANS solver; shaded ellipses: extent of each cluster.}
    \label{fig:app_elastic_net_centroids}
\end{figure}

\begin{figure}[!htbp]
    \centering
    \includegraphics[width=\linewidth]{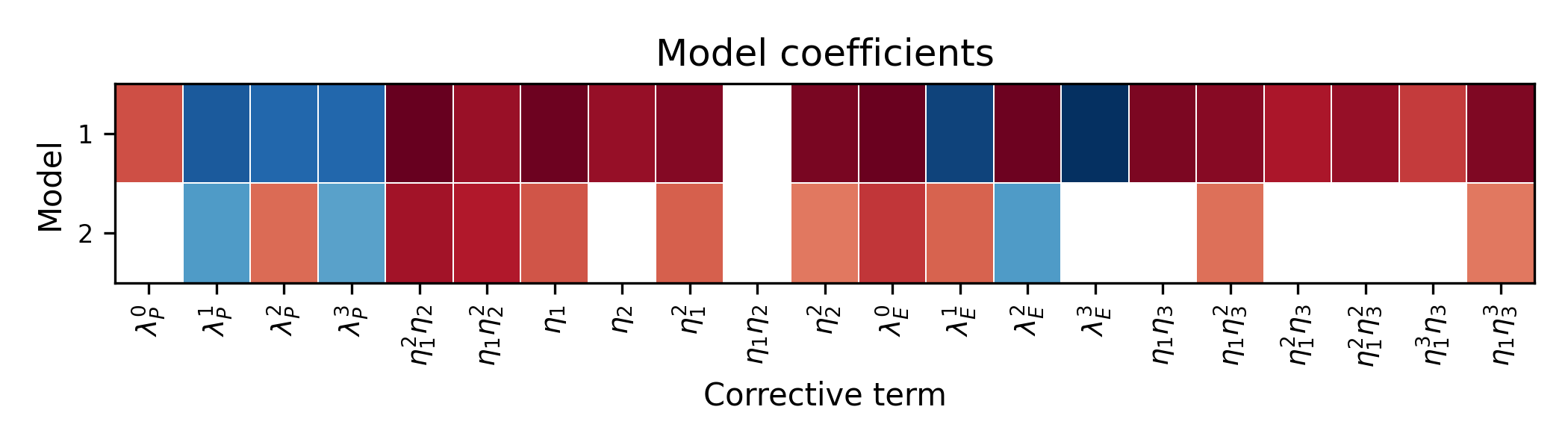}
    \caption{Coefficients of the two cluster centroids propagated through the RANS solver, $M_1^{\mathrm{ZPG}}$ (model 1) and $M_2^{\mathrm{ZPG}}$ (model 2);
    color code as in Fig.~\ref{fig:app_elastic_net_all_models}.}
    \label{fig:app_elastic_net_selected_models}
\end{figure}

\begin{figure}[!htbp]
    \centering
    \begin{subfigure}{0.45\linewidth}
        \centering
        \includegraphics[width=\linewidth]{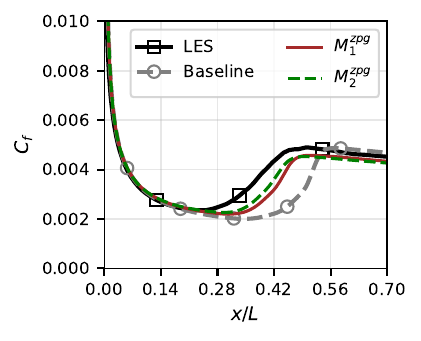}
        \caption{T3A}
        \label{fig:app_elastic_net_t3a_cf}
    \end{subfigure}\hfill
    \begin{subfigure}{0.45\linewidth}
        \centering
        \includegraphics[width=\linewidth]{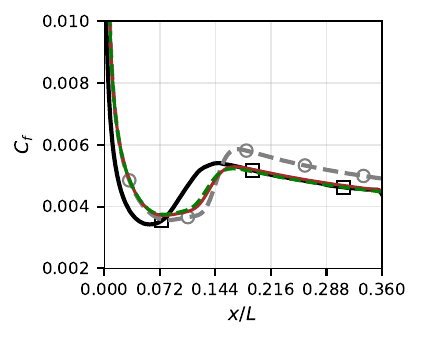}
        \caption{T3B}
        \label{fig:app_elastic_net_t3b_cf}
    \end{subfigure}
    \caption{Cross-validation of the two cluster centroids of the ZPG grid search: skin-friction coefficient predicted on T3A and T3B by $M_1^{\mathrm{ZPG}}$
    (red solid line) and $M_2^{\mathrm{ZPG}}$ (green dashed line), compared with the LES (black line with squares) and the baseline model (gray dashed line
    with circles).}
    \label{fig:app_elastic_net_cf}
\end{figure}

The procedure is illustrated in Figs.~\ref{fig:app_elastic_net_all_models}--\ref{fig:app_elastic_net_cf} for the grid search performed on the two ZPG training
cases, T3A and T3B. Most of the 72 candidates share almost the same dense structure and similar coefficient values, whereas the others are much sparser, and 14 of
them reduce to the baseline model, all their coefficients being pruned (Fig.~\ref{fig:app_elastic_net_all_models}). The silhouette score and the elbow criterion
lead to two clusters, well separated along the first principal component of the descriptors (Fig.~\ref{fig:app_elastic_net_centroids}), so that only two models
instead of 72 have to be propagated through the solver. The centroid of the first cluster, $M_1^{\mathrm{ZPG}}$, retains 20 of the 21 candidate terms, only the
$\eta_1 \eta_2$ monomial of the production exponent being pruned, whereas the centroid of the second cluster, $M_2^{\mathrm{ZPG}}$, is much sparser and retains 13
terms (Fig.~\ref{fig:app_elastic_net_selected_models}).

The skin-friction predictions of the two centroids are compared in Fig.~\ref{fig:app_elastic_net_cf}. On T3A, both move the transition onset upstream, toward the
LES, $M_2^{\mathrm{ZPG}}$ being the closest. On T3B, the two centroids are almost indistinguishable: they slightly reduce the skin-friction overshoot of the
baseline, but the transition onset remains downstream of the LES. The improvement scores confirm this ranking: $M_1^{\mathrm{ZPG}}$ improves on the baseline by
$25.35\%$ on T3A and $2.3\%$ on T3B, i.e.\ $13.82\%$ on average, against $28.14\%$ and $2.6\%$, i.e.\ an average of $15.37\%$, for $M_2^{\mathrm{ZPG}}$, which is therefore
retained for the ZPG cases. The same procedure is applied to the candidates trained on the four training cases, and the retained centroid is the model $M_{G.S}$,
written explicitly in Eq.~\eqref{eq:model_gs}.

%
\section{Leave-one-case-out cross-validation}\label{sec:appendix_lococv}
This appendix provides additional information on the LOCOCV procedure adopted in the bi-level optimization strategy. The four training cases are split into four
subsets (Table~\ref{tab:leave_one_out_subsampling}), each comprising three training cases and one validation case. The bi-level optimization is performed
independently on each subset, which yields four candidate models. For each subset, the leader problem is solved by differential evolution with a population of 40
individuals and a maximum of 100 generations, and each follower problem with a population of 300 individuals and a maximum of 1000 generations. Both use the same
$10^{-5}$ threshold on the loss and are stopped when the best individual has not improved over 20 consecutive generations. Most runs converge well before the maximum
budget, the leader problem requiring 40 generations on average, so that the whole optimization can be completed within a reasonable time on a standard workstation.

Table~\ref{tab:leave_one_out_subsampling} also reports the improvement scores of the four models, each evaluated a posteriori with the RANS solver on its own
validation case. All four models improve over the baseline on the case they have not seen, although only marginally for $M_{S^{(2)}}$ and $M_{S^{(3)}}$, validated
on T3B and T3C3, respectively. The model $M_{S^{(1)}}$, trained on T3B, T3C3 and T3C5, achieves the highest score on its validation case, T3A, and is retained as the
bi-level model $M_{B.L}$, written explicitly in Eq.~\eqref{eq:model_bl}.

\begin{table}[!htbp]
    \centering
    \footnotesize
    \setlength{\tabcolsep}{4pt}
    \begin{tabular}{lcccc}
        \toprule
        Subset                                        & $S^{(1)}$          & $S^{(2)}$         & $S^{(3)}$         & $S^{(4)}$       \\
        \midrule
        Training cases & \makecell{T3B, T3C3,\\ T3C5} & \makecell{T3A, T3C3,\\ T3C5} & \makecell{T3A, T3B,\\ T3C5} & \makecell{T3A, T3B,\\ T3C3} \\
        Validation case                               & T3A                & T3B               & T3C3              & T3C5            \\
        \midrule
        $m_{\overline{u}}$ (\%)                       &  27.15              &   4.09             & 1.3              & 20.4           \\
        $m_{\partial \overline{u} / \partial y}$ (\%) &  37.81             &   5.78             & 2.4              & 33.97           \\
        $m_{C_f}$ (\%)                                &  24.82             &   2.59             & 1.14             & 23.69           \\
        $m_{\mathrm{model}}$ (\%)                     &  29.92             &   4.15             & 1.61             & 26.02           \\
        \bottomrule
    \end{tabular}
    \caption{Leave-one-case-out splits of the four training cases used in the bi-level optimization, and improvement scores of the resulting models
    $M_{S^{(k)}}$, each evaluated with the RANS solver on its own validation case. The model obtained with subset $S^{(1)}$ is retained as $M_{B.L}$.}
    \label{tab:leave_one_out_subsampling}
\end{table}

\end{appendices}

\end{document}